%% file: PRD-Ver-2.tex
\documentclass[preprint,aps,prd,showpacs,nofootinbib,superscriptaddress,floatfix,showkeys]{revtex4-2}

\usepackage{amsfonts}
\usepackage{amsmath}
\usepackage{amssymb} 
\usepackage{bm} % For bold math (esp of lower greek letters)
\usepackage{color}
\usepackage{enumerate}
\usepackage{graphicx}  % Needs to come after graphicx
\usepackage{dcolumn}
\usepackage{hyperref}
\usepackage{lineno}
\hypersetup{
	colorlinks=true,       % false: boxed links; true: colored links
	linkcolor=blue,          % color of internal links
	citecolor=magenta,        % color of links to bibliography
	filecolor=cyan,      % color of file links
	urlcolor=magenta           % color of external links
}

\labelformat{section}{#1} 
\labelformat{subsection}{Section #1} 
\labelformat{figure}{#1} 
\labelformat{subfigure}{Fig.~\thefigure#1} 
\labelformat{table}{Tab.~#1} 

\def\be{\begin{equation}}
	\def\ee{\end{equation}}
\def\ben{\begin{eqnarray}}
	\def\een{\end{eqnarray}}

\begin{document}
\title{Excursion Set Approach to Primordial Black Holes: Cloud-in-Cloud and Mass Function Revisited} 
%\author{xyz}
%\author{xyz}
\author{Ashu Kushwaha} 
\email{kushwaha.a.aa@m.titech.ac.jp}
\author{Teruaki Suyama}
\email{suyama@phys.sci.isct.ac.jp}
\affiliation{Department of Physics, Institute of Science Tokyo, 2-12-1 Ookayama, Meguro-ku,
	Tokyo 152-8551, Japan}
%
%	\author{}
%	\email{}
%	\affiliation{}
	%
	%
   % \date{\today}
\begin{abstract}
The abundance and mass function of primordial black holes (PBHs) are often estimated using the Press–Schechter (PS) formalism. In the case of halo formation, the PS formalism suffers from the miscounting of regions collapsing into halos, known as the cloud-in-cloud problem, which is usually corrected by introducing a multiplicative `fudge factor 2'. By analogy, this factor has sometimes been applied to PBH calculations, although its validity has remained unsettled.
We reformulate the PS approach for PBHs (forming during radiation-dominated epoch) within the excursion-set framework, where the smoothed density contrast undergoes a stochastic random walk as the smoothing scale varies and collapse is identified with the first threshold crossing. While the halo case is described by a Markovian process, we show that the PBH case is non-Markovian, even when the sharp-k filter Window function is adopted. Decomposing the total collapse probability into two distinct components of the stochastic motion, we numerically confirm that their contributions are exactly equal in the case of halo formation, justifying the fudge factor. For PBHs, however, we demonstrate that this equality no longer holds, and consistent inclusion of both contributions is essential to ensure a positive-definite mass function. Our results clarify the origin of the ambiguity surrounding the fudge factor and establish a robust theoretical foundation for PBH abundance calculations.

\end{abstract}
\maketitle
\newpage

\section{Introduction}

Our understanding of the formation of observed structures in the universe, such as stars, galaxies, galaxy clusters, and dark matter halos, is based on the fact that initial overdense regions with small amplitude grow and gravitationally collapse to form these virialized objects (halos). 
In 1974, Press \& Schechter (PS) proposed the analytic framework to calculate the mass distribution (spectrum) of these virialized objects by assuming Gaussian random density fluctuations and spherical collapse model~\cite{Press:1973iz}. 
Furthermore, they provided an interesting connection between the statistical properties of initial linear density fluctuations to the eventual abundance of collapsed objects, which are non-linear structures. The PS formalism suggests that the regions within the universe that possess an overdensity exceeding a specific critical threshold value $\delta_c$ will undergo gravitational collapse to form virialized dark matter halos. However, the formula derived by PS predicted that only half of the matter in the initial density field would collapse into halos. 
To restore mass conservation, they introduced a multiplication by a factor of two (the so-called `fudge factor 2').
This `fudge factor 2' is related to the so-called cloud-in-cloud problem, which leads to the miscounting of the number of regions collapsing to halos. The ``cloud-in-cloud" problem arises when small underdense regions are embedded in larger overdense regions which become part of bigger halos, or small overdense regions are embedded in a bigger overdense region, both leading to miscounting the number of regions collapsing to halos.
This incorrect counting led the prediction of PS formalism to account for only half of the mass fraction. 

Bond et al.~\cite{Bond:1990iw}, proposed a rigorous and statistically robust framework for the solution to the cloud-in-cloud problem based on the theory of excursion set of density perturbation field $\delta({\bm x},R_M)$ smoothed with a continuous hierarchy of filters of radii $R_M$. Interestingly, this approach naturally accounts for the ``fudge factor 2" without any ad-hoc multiplication.
In this theory, in sharp $k-$ space filtering, the evolution of density perturbations of a fixed spatial point ${\bm x}$ i.e. $\delta (R_M)$  can be modeled as a stochastic Brownian random walk with respect to smoothing scale\footnote{If the steps of this random walk are uncorrelated, it is referred to as a Markov process 
	while the process is called non-Markov if the steps at different times are correlated. The latter may arise 
	when the density perturbation is smoothed with filters such as the top-hat
	filter in coordinate space, or when one considers
	non-Gaussian fluctuations, see Refs.\cite{Maggiore:2009rv,Maggiore:2009rw,Maggiore:2009rx}.}.
Within this framework, halo formation is understood to occur when this random walk crosses a critical threshold barrier $\delta_c$. Therefore, the excursion set theory provides a more comprehensive framework by tracking the evolution of density perturbations crossing the barrier $\delta (R_M) > \delta_c$, which lead to collapse to form halos~\cite{Zhang:2005ar,Nikakhtar:2018qqg,Baghram:2019jlu}. 
Note that alternative approaches to the cloud-in-cloud problem are also proposed in Ref.~\cite{Peacock:1990zz,Jedamzik:1994nr}.

Primordial black holes (PBHs) with mass larger than $10^{15}\, {\rm g}$ are considered interesting candidates for dark matter. Despite the absence of direct observational evidence of PBHs, there are various indirect hints of their existence~\cite{2018-Sasaki.etal-CQG,2020-Carr.etal-Rept.Prog.Phys}. However, after the discovery of gravitational waves, PBHs have received renewed interest, as the event GW150914, observed by LIGO~\cite{LIGOScientific:2016aoc}, can be explained by the coalescence of PBHs~\cite{Sasaki:2016jop,Bird:2016dcv}.
PBHs are formed from the collapse of large-amplitude small-scale density fluctuations during horizon re-entry in the early universe, for example, during the radiation-dominated (RD) epoch. 
To achieve that, the amplitude of the primordial power spectrum has to be amplified by about $\mathcal{O}(10^7)$ of magnitude above its amplitude observed on the cosmic microwave background scale (i.e.,$\mathcal{P}_{\zeta}\sim 2\times 10^{-9}$), which can be achieved by introducing nontrivial physics (such as features in the inflationary potential etc); see Refs.~\cite{Garcia-Bellido:1996mdl,Yokoyama:1998pt,Suyama:2014vga,Nakama:2016kfq,Garcia-Bellido:2017mdw,Inomata:2018cht,Bhaumik:2019tvl,Braglia:2020eai,Kawasaki:2012wr}.
Abundance and mass function of PBHs have been mostly discussed within the context of two approaches~\cite{Young:2014ana,2018-Sasaki.etal-CQG,Yoo:2020dkz,Young:2020xmk}: Press-Schechter formalism or Peak theory. However, the estimation of the PBH mass distribution (or the abundance of PBHs) differs between the two approaches, and the preference for one over the other is still unsettled.

The excursion set formalism is a powerful tool to systematically calculate the mass function, the clustering properties of dark matter halos relative to the dark matter, and it can be applied to numerous other problems~\cite{Paranjape:2011wa,Paranjape:2012jt,Paranjape:2012ks,Musso:2013pha,Musso:2013pja,Nikakhtar:2018qqg,Baghram:2019jlu,Auclair:2024jwj}. The excusrion set formalism to PBHs formation during matter-domination epoch is discussed in Ref.\cite{Auclair:2020csm}. In this work, we use this approach to gain insight into the cloud-in-cloud problem for PBHs formation during RD era in PS formalism. In particular, we investigate whether the `fudge factor 2' is important in the calculation of PBH mass fraction.
There has been a confusion in the literature about using/not using this factor in the formula to calculate the mass fraction of PBHs, see for example Ref.\cite{Garcia-Bellido:1996mdl,Yokoyama:1998pt,Garcia-Bellido:2017mdw,Inomata:2018cht,Bhaumik:2019tvl,Braglia:2020eai,Kawasaki:2012wr,Kushwaha:2024zhd}. However, to the best of our knowledge, no attempt have been made to resolve this issue, and in this work we investigate this in detail\footnote{`PBHs in the excursion set theory' has been studied in Refs.\cite{Erfani:2021rmw,Kameli:2025qzp}, however, the main focus of these works is different from ours and does not address the issues discussed in our work.}. We also show how the naive calculations often lead to a negative mass function in some mass ranges and provide a robust theoretical framework for PBH abundance calculations. The paper is organised as follows. First, we review the application of excursion set theory by considering numerous examples that demonstrate how it solves the cloud-in-cloud problem, which justifies the `fudge factor 2' for the halo formation case. Then, we examine the formation of PBHs within the PS formalism and discuss the irrelevance of `fudge factor 2' and the issues related to the mass function calculations.

\section{A brief review on explanation of the fudge factor 2 in the PS formalism}
A basic assumption in the original Press-Schechter formalism is that an overdense region,
when the smoothed density contrast exceeds a threshold $\delta_c$, collapses to form a gravitationally
bound system called halo.
More precisely, the point ${\bm x}$ is contained in a halo with mass greater than $M$
if $\delta ({\bm x};R_M)$ defined by
\be
\delta ({\bm x};R_M) = \int d{\bm x'} ~W({\bm x'}-{\bm x};R_M) \delta ({\bm x'}),
\ee 
which represents the density contrast smoothed over the comoving scale $R_M$
corresponding to the mass $M$\footnote{Relation
	between $R_M$ and $M$ is given by $M=\frac{4\pi}{3}\rho_{m,0} R_M^3$, where $\rho_{m,0}$ is the present matter density.},
is larger than $\delta_c$.
Here $W({\bm x};R_M)$ is the Window function which smoothes out the structure
smaller than the length scale $R_M$.
In our study, we adopt the sharp-k Window function for which
the Window function in the Fourier space becomes \footnote{
	The Fourier transform of $W$ is defined by
	\be
	{\tilde W} (k;R_M)=\int d{\bm x}~e^{-i{\bm k}\cdot {\bm x}} W ({\bm x};R_M).
	\ee
}
\be
{\tilde W}(k;R_M)=\Theta ( 1-kR_M)~,
\ee
where $\Theta$ is the Heaviside function.
As it is clear from this expression, this Window function sharply
eliminates the modes whose wavenumber is greater than $R_M^{-1}$.

The variance of $\delta ({\bm x};R_M)$ is given by
\be
\sigma_M^2 =\langle \delta^2 ({\bm x};R_M) \rangle
=\int \frac{d{\bm k}}{{(2\pi)}^3} P_\delta (k) {\tilde W}^2 (k;R_M)
=\int^{k<R_M^{-1}} \frac{d{\bm k}}{{(2\pi)}^3} P_\delta (k),
\ee
where $P_\delta (k)$ is the power spectrum of $\delta$ defined by
\be
\langle {\tilde \delta} ({\bm k}) {\tilde \delta} ({\bm k'}) \rangle =
{(2\pi)}^3 P_\delta (k) \delta ({\bm k}+{\bm k'}),
\ee
and ${\tilde \delta}$ is Fourier transform of $\delta$:
\be
{\tilde \delta} ({\bm k})=\int d{\bm x}~e^{-i{\bm k}\cdot {\bm x}} \delta ({\bm x}).
\ee
Because the upper limit of integration is a monotonically increasing function 
of $M^{-1}$,
$\sigma_M^2$ also monotonically increases as $M^{-1}$ is increased.

Now, let us monitor how $\delta ({\bm x};R_M)$ varies as the smoothing scale is changed.
Introducing an effective time $\tau$ by $\tau=\frac{1}{R_M}$, we can regard the variation
of $\delta ({\bm x}; \tau)$ at a fixed point ${\bm x}$ as a time evolution of a Brownian motion of a 
point particle such that the initial condition is $\delta (\tau=0)=0$ and 
the two-point function is given by
\begin{align}
	\langle \delta (\tau) \delta (\tau') \rangle= &
	\int \frac{d{\bm k}}{{(2\pi)}^3} {\tilde W} (k;\tau^{-1}) {\tilde W}(k;\tau'^{-1})
	P_\delta (k) \nonumber \\
	&=\int \frac{dk}{k} {\cal P}_\delta (k) 
	\Theta (1-\frac{k}{\tau}) \Theta (1-\frac{k}{\tau'})~,
\end{align}
where we used ${\cal P}_\delta =\frac{k^3}{2\pi^2} P_\delta$.
This statistical property is derived by modeling that the equation of motion
of the point particle is given by the following Langevin equation;
\be
\label{Langevin-eq}
\frac{d}{d\tau} \delta (\tau)=\xi (\tau),
\ee
where $\xi (\tau)$ is a Gaussian noise with the following statistical properties~\cite{RevModPhys.15.1}:
\be
\label{xi-stat-properties}
\langle \xi (\tau) \rangle =0,~~~~~\langle \xi (\tau) \xi (\tau')\rangle=C(\tau, \tau').
\ee
Here the two-point function of $\xi$ is related to that of $\delta (\tau)$ by
\be
\label{PS-C-delta}
C(\tau, \tau')=\frac{\partial^2}{\partial \tau \partial \tau'}
\langle \delta(\tau) \delta (\tau') \rangle = \frac{1}{\tau}
{\cal P}_\delta (\tau) \delta (\tau-\tau').
\ee
Notice that this is a positive-definite matrix in the sense that
\be
\int d\tau \int d\tau' ~z(\tau) z(\tau')C(\tau, \tau')
=\bigg\langle {\left( \int d\tau ~z(\tau) \frac{d}{d\tau}\delta( \tau) \right)}^2 \bigg\rangle \ge 0,
\ee
for any function $z(\tau)$,
providing an explicit confirmation of the property the two-point function 
of the noise $\xi$ must satisfy.
\begin{figure}[t]
	\begin{center}
		\includegraphics[clip,width=11.0cm]{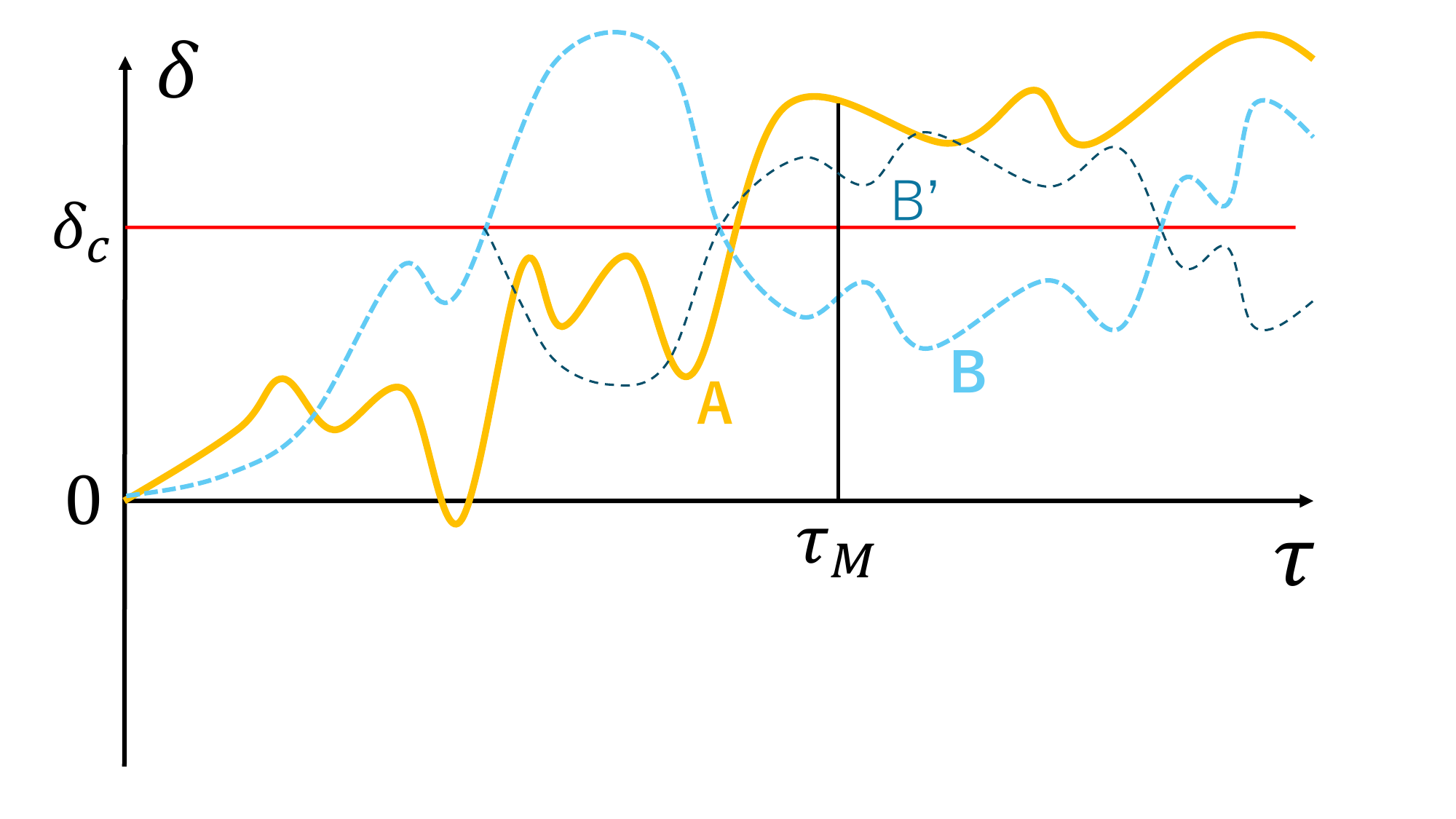}
		\caption{Schematic picture showing the variation of the smoothed density contrast $\delta (\tau)$
			for three cases. At a specific scale $\tau_M$, a region corresponding to the trajectory A is in
			a halo with mass larger than $M$. A region corresponding to the trajectory B is $\delta < \delta_c$ but is still 
			part of a halo with mass larger than $M$ because $\delta (\tau)$ exceeds $\delta_c$ for some values $\tau< \tau_M$.}
		\label{fig:PS-fig1}
	\end{center}
\end{figure}

Fig.~\ref{fig:PS-fig1} provides a schematic picture.
At a specific scale $\tau_M$, the region corresponding to the trajectory A is in 
a halo with mass larger than $M$ because $\delta (\tau_M) > \delta_c$.
Any trajectories satisfying $\delta (\tau_M) >\delta_c$ represent the situation that the
location under consideration is inside the halo with mass $>M$.
Additionally, even if $\delta (\tau_M) <\delta_c$, the region is still a part of the
halo with mass $>M$ if there exists $\tau$ such that $\tau<\tau_M$ and $\delta (\tau) > \delta_c$
as it is represented as the trajectory B.
From this consideration, it can be said that the probability that a given point ${\bm x}$ is in
part of a halo with mass $>M$ is 
\be
\label{fraction:PS-formalism}
P(>M) =P(\delta(\tau_M)>\delta_c) + P(\delta (\tau_M) < \delta_c \cap
\exists\, \tau < \tau_M \;\; \text{s.t.} \;\; \delta( \tau ) > \delta_c ).
\ee
The first/(second) term on the right-hand side represents the probability of having the trajectory like A/(B). For notational simplicity, in what follows,
we denote the first and second terms on the right-hand side of Eq.~(\ref{fraction:PS-formalism}) as $P_1$ and $P_2$, respectively.
The first term is simply given by
\be
\label{P:first-term}
P_1 (\tau_M) \equiv P(\delta(\tau_M)>\delta_c) =\int_{\delta_c}^\infty \frac{1}{\sqrt{2\pi} \sigma (\tau_M)}
\exp \left( -\frac{\delta^2}{2\sigma^2 (\tau_M)} \right) d\delta.
\ee

\begin{figure}[t]
	\begin{center}
		\includegraphics[clip,width=7.0cm]{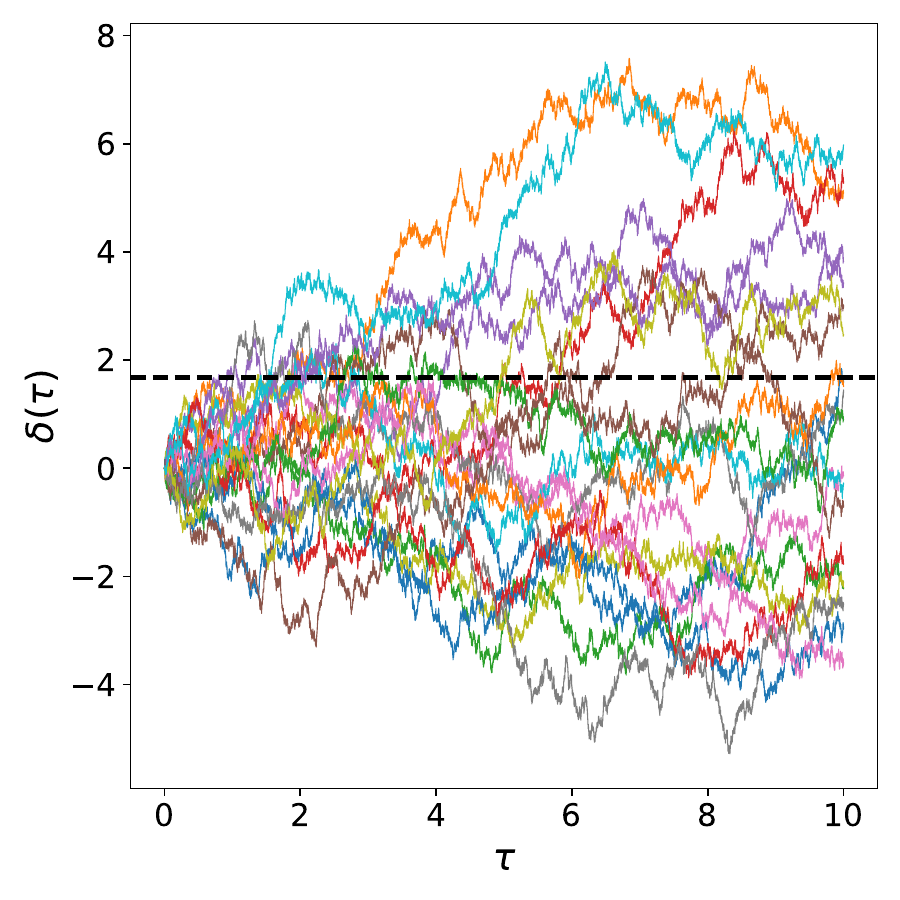}
		\includegraphics[clip,width=7.0cm]{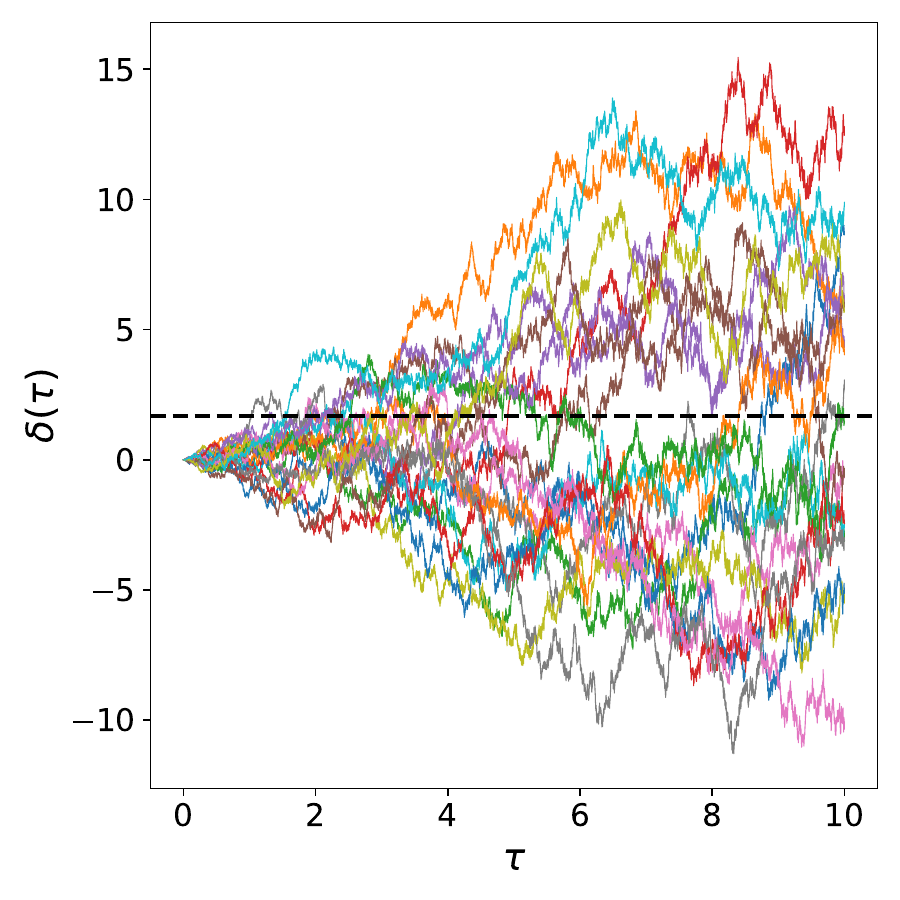}
		\caption{Showing the variation of smoothed density contrast $\delta (\tau )$ (Brownian motion) for the $20$ Langevin trajectories for power law power spectrum~\eqref{power-law-spectrum} for $n=1$ (left-plot) and $n=2$ (right-plot). We set $ \mathcal{A}=1$ and the horizontal dashed line shows the typical value of critical density contrast $\delta_c=1.68$.
		}
		\label{fig:powerlaw_n=1-2}
	\end{center}
\end{figure}

To compute the second term $P_2$, we note that because two $\xi$ at different times
are uncorrelated (see Eq.~(\ref{PS-C-delta})) the trajectory B is as equally probable as
the trajectory B' which is obtained by reflecting the trajectory B after it hits 
$\delta_c$ for the first time;
\be
\delta_{B'}(\tau)=
\left\{
\begin{array}{l}
	\delta_B (\tau)  ~~~~\tau<\tau_{\rm fa}\\
	2\delta_c-\delta_B (\tau) ~~~~\tau\ge \tau_{\rm fa},
\end{array}
\right.
\ee
where $\tau_{\rm fa}$ is the first-arrival time.
Then the second term is equal to the probability of having the trajectories like B'
and it is equal to
$P_1$.
Thus, Eq.~(\ref{fraction:PS-formalism}) becomes
\be
P(>M) =2P_1 (\tau_M)=2\int_{\delta_c}^\infty \frac{1}{\sqrt{2\pi} \sigma (\tau_M)}
\exp \left( -\frac{\delta^2}{2\sigma^2 (\tau_M)} \right) d\delta.
\ee
In this way, the famous `fudge factor 2' has been obtained.

\subsection{Numerical Simulations}

\begin{figure}[t]
	\begin{center}
		\includegraphics[clip,width=7.0cm]{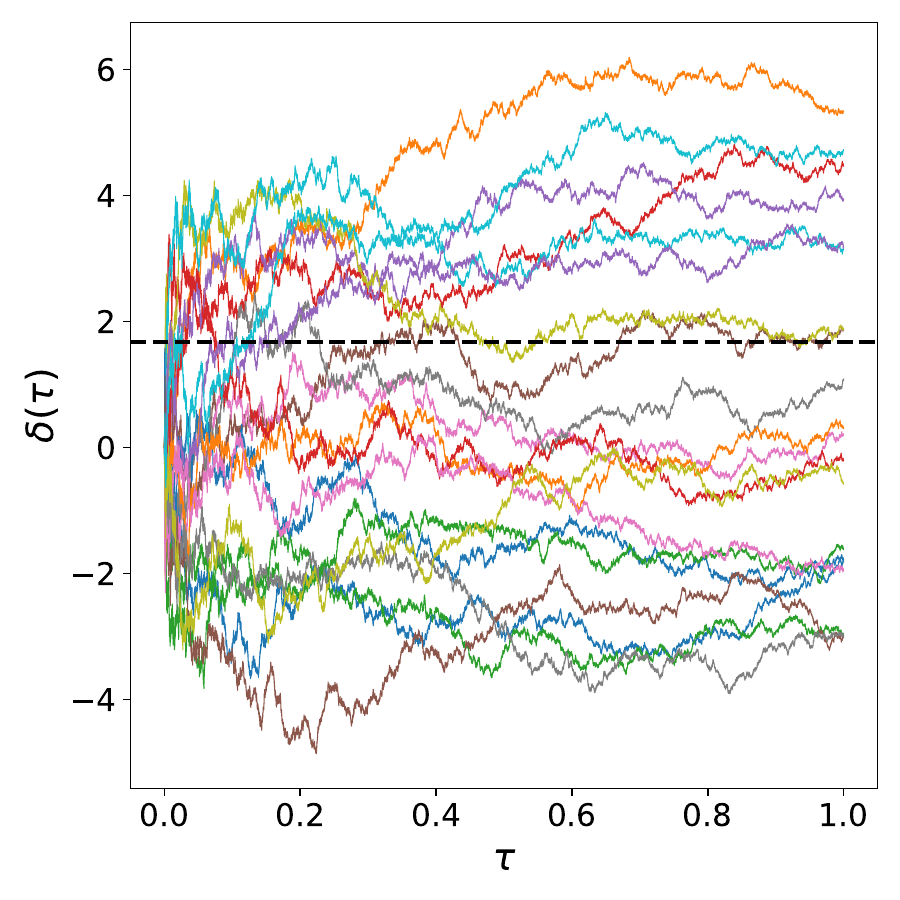}
		\includegraphics[clip,width=7.0cm]{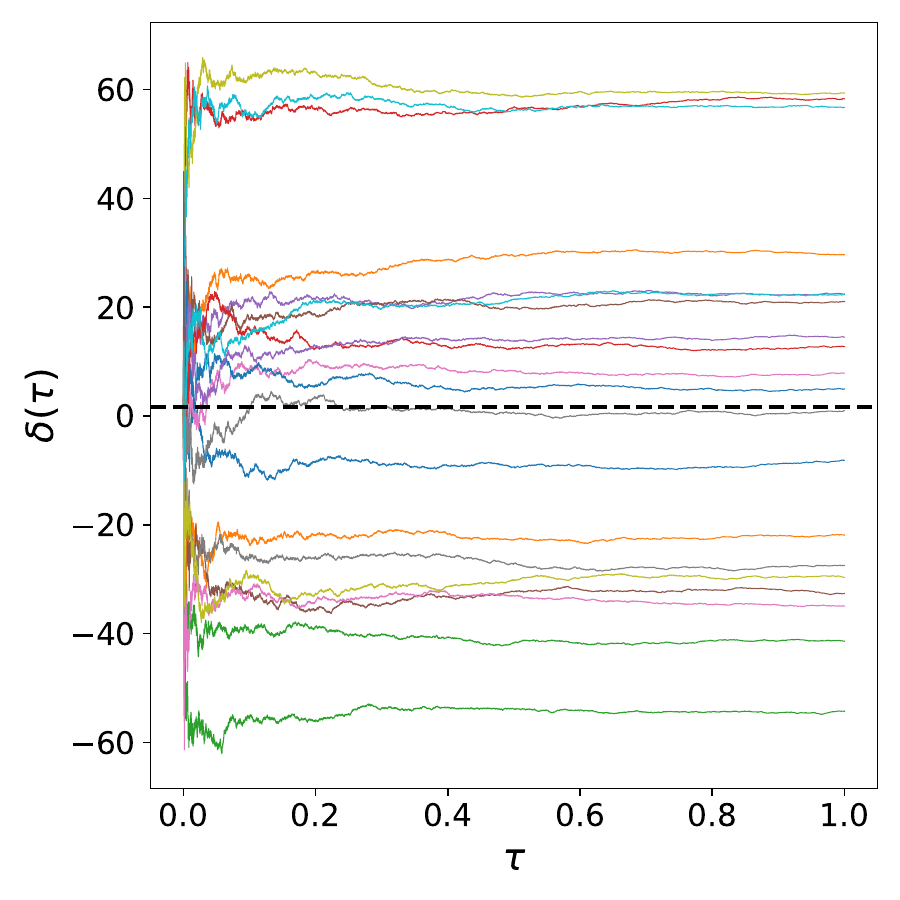}
		\caption{Showing the variation of smoothed density contrast $\delta (\tau )$ (Brownian motion) for the $20$ Langevin trajectories for power law power spectrum~\eqref{power-law-spectrum} for $n=0$ (left-plot) and $n=-1$ (right-plot). We set $ \mathcal{A}=1$ and the horizontal dashed line shows the typical value of critical density contrast $\delta_c=1.68$. The divergence of the power spectrum at $\tau \rightarrow0$ is regularized by introducing small $\epsilon=10^{-3}$ as $\tau \rightarrow (\tau +\epsilon)$ with $\epsilon \gg$ time-step size used in the simulations.
		}
		\label{fig:powerlaw_n_regularized}
	\end{center}
\end{figure}

After discussing the method to obtain the `fudge factor 2', which is based on the theory of excursion set, let us now justify the conclusion by considering some examples of power spectrum $\mathcal{P}_{\delta} (\tau)$ in Eq.\eqref{PS-C-delta}. Specifically, we simulate the Langevin equation~\eqref{Langevin-eq} with the covariance matrix given by Eq.\eqref{PS-C-delta}. 
As mentioned before, because 
$C (\tau,\tau') \propto \delta (\tau-\tau')$, the $\xi (\tau)$ represents 
an uncorrelated
Gaussian noise (i.e., white noise).

\begin{figure}[t]
	\begin{center}
		\includegraphics[clip,width=7.0cm]{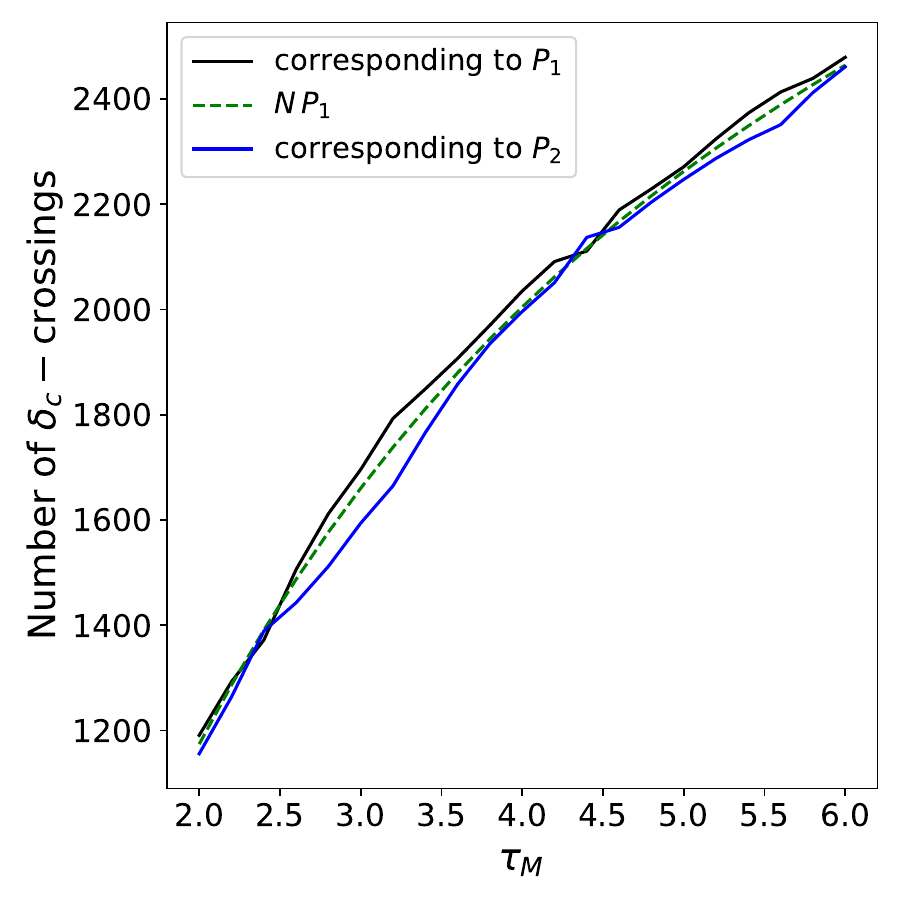}
		\includegraphics[clip,width=7.0cm]{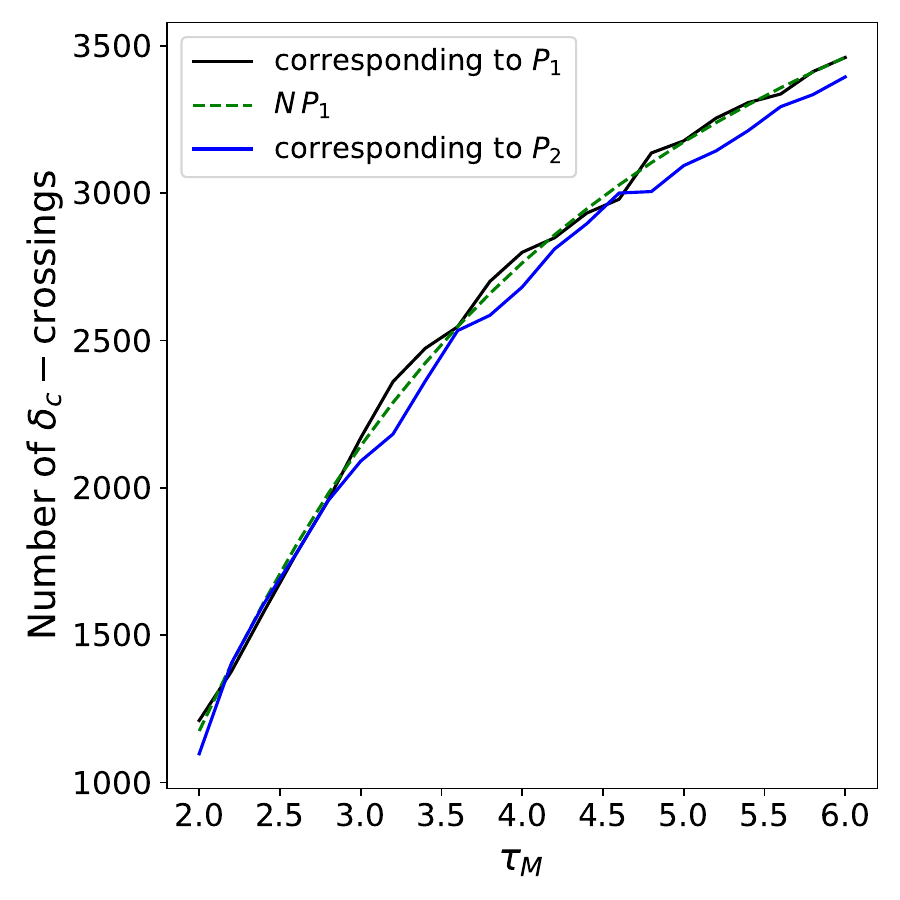}
		\caption{
			Left panel: The black and blue curves represent the number of trajectories computed from $N=10^4$ realizations, corresponding to $P_1$ and $P_2$, respectively, assuming the power-law power spectrum~\eqref{power-law-spectrum} with $n=1$.
			The green dotted curve shows $N P_1 (\tau_M)$ computed from Eq.~(\ref{P:first-term}) which is the expected number of trajectories corresponding to $P_1$.
			The agreement between the black curve and the green dotted curve, 
			within statistical uncertainty ($\sim \sqrt{N}$), indicates that the simulations were performed correctly.
			Right panel: Same as the left panel but for $n=2$.
		}
		\label{fig:powerlaw_n=1-2_path_crossing}
	\end{center}
\end{figure}

\begin{figure}[h]
	\begin{center}
		\includegraphics[clip,width=7.0cm]{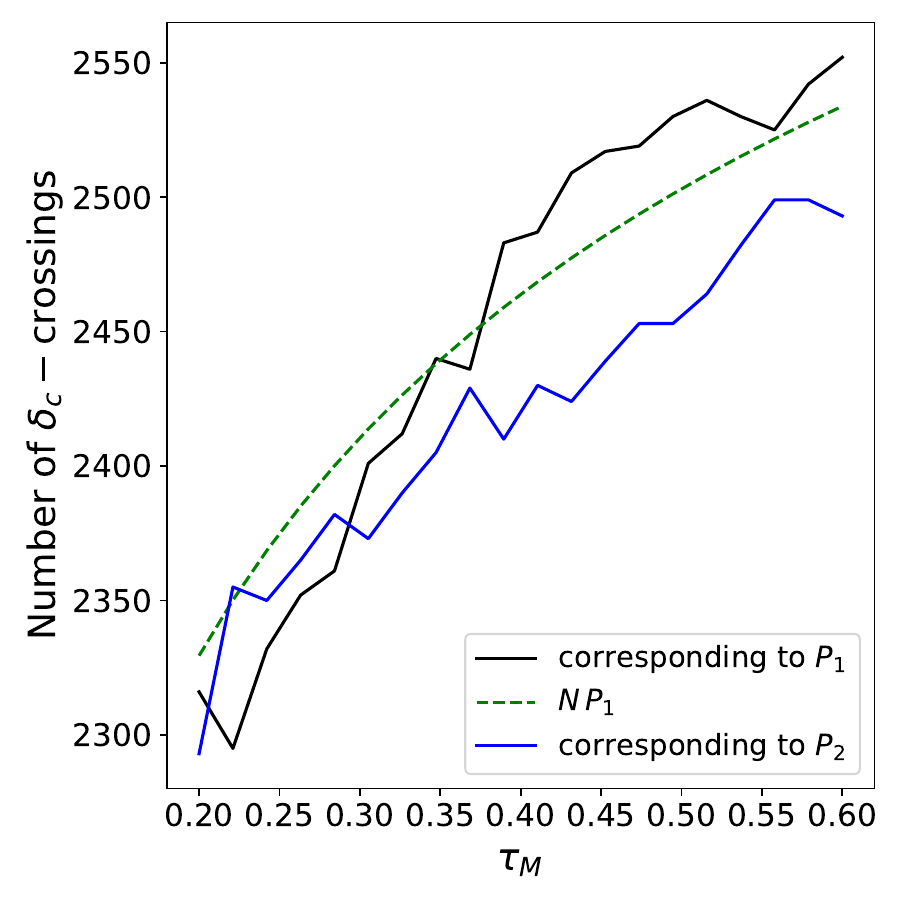}
		\includegraphics[clip,width=7.0cm]{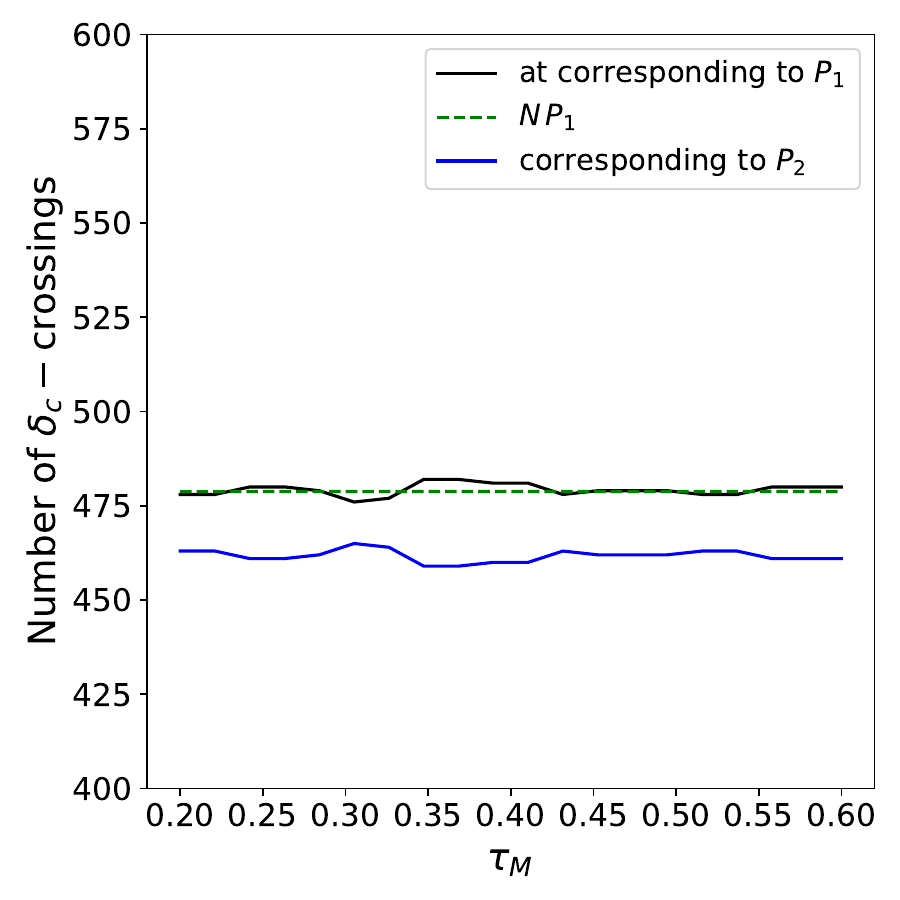}
		\caption{Left panel: The black and blue curves represent the number of trajectories computed from $N=10^4$ realizations, corresponding to $P_1$ and $P_2$, respectively, assuming the power-law power spectrum~\eqref{power-law-spectrum} with $n=0$.
			The green dotted curve shows $N P_1 (\tau_M)$ computed from Eq.~(\ref{P:first-term}) which is the expected number of trajectories corresponding to $P_1$.
			The agreement between the black curve and the green dotted curve, 
			within statistical uncertainty ($\sim \sqrt{N}$), indicates that the simulations were performed correctly.
			Right panel: Same as the left panel but for $n=-1$ and $N=10^3$ realizations. 
			The divergence of the power spectrum at $\tau \rightarrow0$ is regularized by introducing small $\epsilon=10^{-3}$.
		}
		\label{fig:powerlaw_n_regularized-cross}
	\end{center}
\end{figure}

First, we consider the power spectrum of the density perturbations as a power law
\begin{align}\label{power-law-spectrum}
	\mathcal{P}_{\delta} (\tau) = \mathcal{A} \tau^n \qquad \text{with} \quad C (\tau,\tau') = \mathcal{A} \tau^{n-1} \delta (\tau-\tau') ~
\end{align}
where $\mathcal{A}$ is the amplitude (which is set to unity in the simulations unless specified) and $n$ is the spectral index. 
For $n\neq1$, we have the $\tau$-dependent white noise, which has an increasing or decreasing variance, depending on the spectral index $n$. 
From Eq.\eqref{power-law-spectrum}, we can see that the variance $\langle \delta (\tau)^2\rangle $ is a well-behaved function of $\tau$ for $n\geq1$ and diverges for $n<1$ as $\tau \rightarrow 0$.
However, the divergence at $\tau \rightarrow0$ can be regularized by introducing a small regularization parameter $\epsilon$, i.e., $\tau \rightarrow \tau + \epsilon$ such that $\epsilon \gg d\tau$, where $d\tau$ is the time-step in the simulations. In this way, the regularized covariance matrix becomes $C_{\rm r} (\tau,\tau') = \mathcal{A} (\tau +\epsilon)^{n-1} \delta (\tau-\tau')$ and avoids the divergence\footnote{ For example, for $n=0$, the variance $\langle \delta (\tau)^2\rangle = \int_0^{\tau} \int_0^{\tau} d\tau' d\tau'' \langle \xi (\tau') \xi (\tau'') \rangle = \int_0^{\tau}  d\tau' \mathcal{A} (\tau' +\epsilon)^{-1} =\mathcal{A} \ln{\left( \frac{\tau+\epsilon}{\epsilon} \right)}$ is well defined as $\tau \rightarrow 0$, and hence there is no logarithmic divergence. Similarly for $n=-1$ $\langle \delta (\tau)^2\rangle = \mathcal{A}  \left( \frac{1}{\epsilon} - \frac{1}{\tau+\epsilon} \right)$ is also well defined as $\tau \rightarrow 0$.
}.
Figure~\ref{fig:powerlaw_n=1-2} shows the Brownian motion of smoothed density contrast $\delta (\tau)$ of the trajectories for $n=1$ (i.e., $\tau-$independent white noise) and $n=2$ (i.e., $\tau-$dependent white noise). 
Similarly, Figure~\ref{fig:powerlaw_n_regularized} shows the Brownian motion of smoothed density contrast $\delta (\tau)$ of the trajectories for $n=0$ and $n=-1$ by using the regularization to avoid the divergence at $\tau\rightarrow0$. 
Both figures show the simulation results for various cases of noise spectrum $\xi(\tau)$ for power law, as illustrated in the schematic diagram of Figure~\ref{fig:PS-fig1}.
Furthermore, to understand the probability that a given point ${\bm x}$ is in part of a halo with mass $>M$, we simulate $10^4$ trajectories with varying $\tau_M$ (which we refer to as \textit{$\delta_c-$crossing statistics plot}) and count the number of trajectories corresponding to 
the first term  
i.e., $P_1$ and the second term $P_2$, in Eq.\eqref{fraction:PS-formalism}. 
Figure~\ref{fig:powerlaw_n=1-2_path_crossing} and \ref{fig:powerlaw_n_regularized-cross} show the $\delta_c$-crossing statistics plots, 
as we can see that both terms are equally probable within the statistical uncertainty. 
Therefore, our simulations demonstrate that trajectories B and B' in Figure~\ref{fig:PS-fig1} are equally probable.

\begin{figure}[t]
	\begin{center}
		\includegraphics[clip,width=7.0cm]{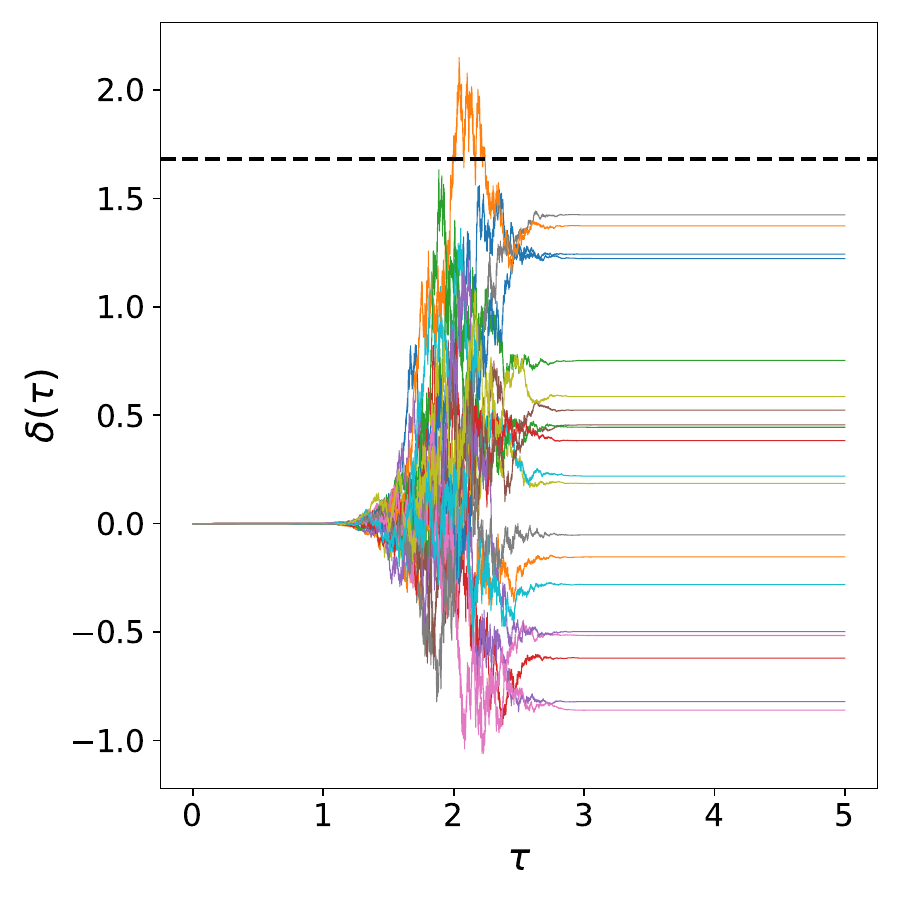}
		\includegraphics[clip,width=7.0cm]{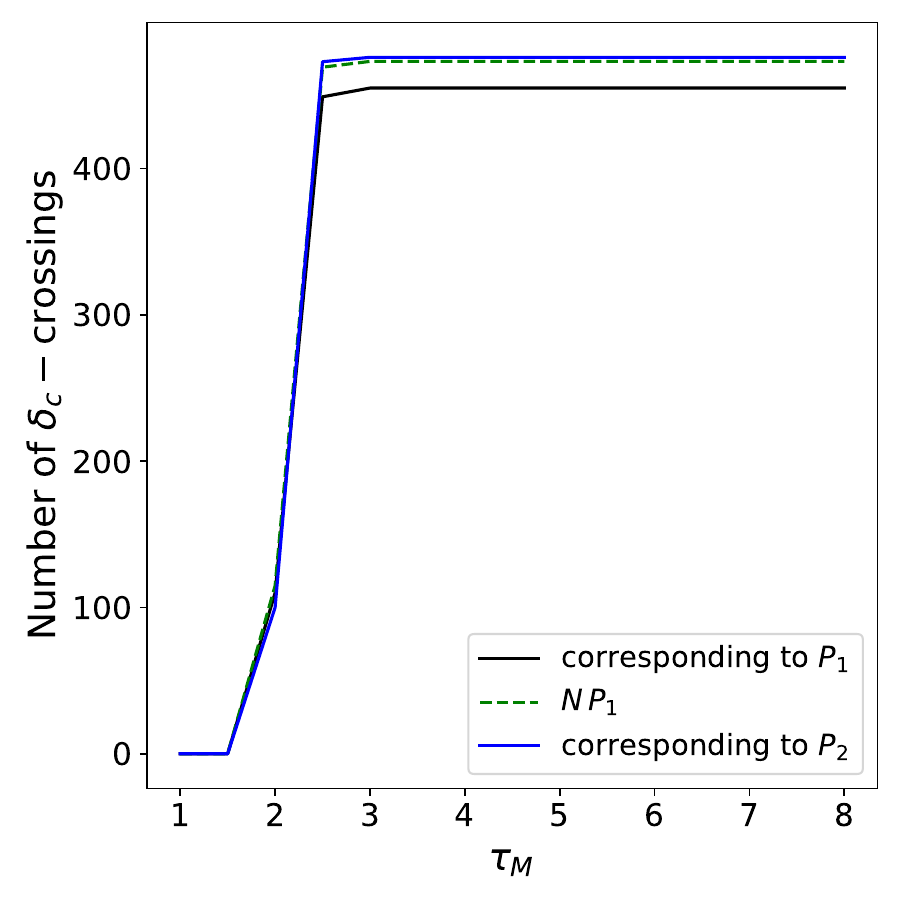}
		\caption{Left panel: Variation of smoothed density contrast $\delta (\tau )$ (Brownian motion) for the $20$ Langevin trajectories for sharply peaked power spectrum~\eqref{spiky-spectrum} for $\tau_*=2, \mathcal{A}=1,\Delta=0.2$ and $\delta_c=1.68$. Right panel: The black and blue curves represent the number of trajectories computed from $N=10^4$ realizations, corresponding to $P_1$ and $P_2$, respectively, for the same sharply peaked power spectrum~\eqref{spiky-spectrum}.
			The green dotted curve shows $N P_1 (\tau_M)$ computed from Eq.~(\ref{P:first-term}) which is the expected number of trajectories corresponding to $P_1$.
		}
		\label{fig:spiky-power-spectrum}
	\end{center}
\end{figure}

Next, we consider a sharply peaked power spectrum of the form
\begin{align}\label{spiky-spectrum}
	\mathcal{P}_{\delta} (\tau) = \mathcal{A} \tau_* \,\frac{1}{\sqrt{2\pi}\Delta} e^{-\frac{(\tau-\tau_*)^2}{2\Delta^2}} 
	\qquad \text{with} \quad C (\tau,\tau') = \frac{\mathcal{A} \tau_*}{\tau} \frac{1}{\sqrt{2\pi}\Delta} e^{-\frac{(\tau-\tau_*)^2}{2\Delta^2}} \delta (\tau-\tau') ~.
\end{align}
where the power spectrum peaks at $\tau=\tau_*$ with peak amplitude $\mathcal{A}$, which is also set to unity in the simulations unless specified. Note that $\Delta$ controls the width of the peak, for example, the limit $\Delta\rightarrow 0$ gives the Dirac-delta function, and $\Delta \sim \mathcal{O}(1)$ corresponds to a broad power spectrum.
Similarly to the power law case, in Figure~\ref{fig:spiky-power-spectrum}, we plot the variation of smooth density contrast $\delta (\tau)$ (left plot) and the $\delta_c$-crossing statistics (right plot). Note that the width of oscillating features (or standard Brownian motion) in the left plot in Figure~\ref{fig:spiky-power-spectrum} is directly proportional to the width $\Delta$, and the limit $\Delta\rightarrow 0$ (which corresponds to the Dirac-delta function) squeezes them to a very narrow width.

Before closing this section, we would like to mention the following: we can understand the schematic diagram in Figure~\ref{fig:PS-fig1} as representing one particular scenario for the Langevin trajectories shown in Figures~\ref{fig:powerlaw_n=1-2},\ref{fig:powerlaw_n_regularized},\ref{fig:spiky-power-spectrum} (left plot). 
Furthermore, from the plots for $\delta_c$-crossing statistics, we can conclude that the first term $P_1$ and the second term $P_2$ in Eq.\eqref{fraction:PS-formalism} are equally probable. 
Therefore, trajectories like B' corresponding to B (as shown in the schematic diagram in Fig.\ref{fig:PS-fig1}) are equally probable. Thus, `fudge factor 2' derived from Eq.\eqref{fraction:PS-formalism} is justified.

\begin{figure}[t]
	\begin{center}
		\includegraphics[clip,width=11.0cm]{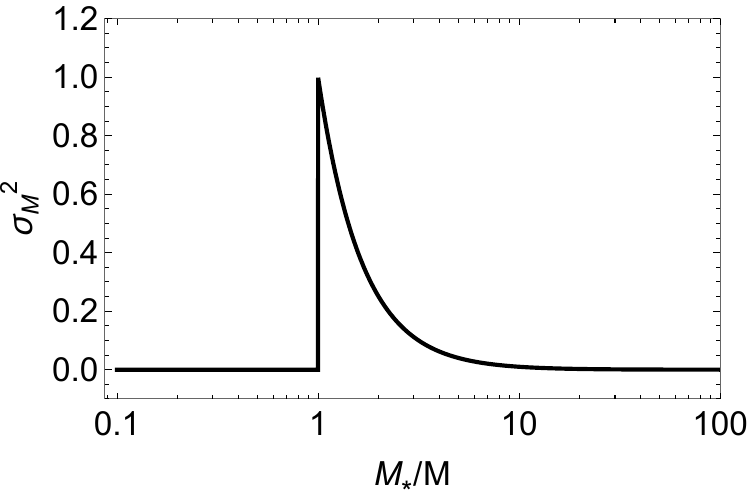}
		\caption{Plot of $\sigma_M^2$ for the sharply peaked spectrum of $\zeta$.}
		\label{fig:sigma-2-PBH}
	\end{center}
\end{figure}

\section{PS formalism in the case of PBH}
Having reviewed the reason for the inclusion of the `fudge factor 2' in the PS formalism,
we now focus on computing the PBH abundance during RD epoch in the framework of the PS formalism.
PBH formation differs from the halo formation in that former occurs on the scale
of the Hubble horizon while the latter occurs on any sub-Hubble scales.
This leads to a natural identification of the PBH mass $M$ with the Hubble horizon by
$M=1/(2GH)$. 
Thus, the density contrast smoothed over the Hubble horizon 
\be
\delta ({\bm x};1/(a_M H_M)) = \int d{\bm x'} ~W({\bm x'}-{\bm x};1/(a_M H_M)) \delta (t_M,{\bm x'}),
\ee
becomes the relevant quantity \footnote{The density contrast is defined to be the
	one on the comoving slice.}, where $a$ and $H$ are scale factor and Hubble parameter, respectively. Note that we use the 
traditional quantity $\delta$, rather than more modern alternatives 
such as the compaction function~\cite{Shibata:1999zs,2019-Musco-PRD,Harada:2023ffo,Harada:2024trx}. 
The specific choice of quantity is not essential to our discussion regarding the fudge factor 2. 
Here the subscript $M$ means that the corresponding quantity is evaluated at the
time when the horizon mass becomes equal to $M$. 
For instance, $H_M=1/(2 GM)$.
The density contrast is related to the curvature perturbation $\zeta$ by
\be
\delta (t,{\bm x})=-\frac{4}{9a^2 H^2} \triangle \zeta ({\bm x}).
\ee
Using this relation, the variance of the smoothed density contrast becomes
\be
\sigma_M^2=\int \frac{d{\bm k}}{{(2\pi)}^3}~ \frac{16}{81} {\left( \frac{k}{k_M} \right)}^4 
\Theta (1-\frac{k}{k_M}) P_\zeta (k)
=\frac{16}{81} \int_0^{k_M} \frac{dk}{k}
{\left( \frac{k}{k_M} \right)}^4 {\cal P}_\zeta (k),
\ee
where $k_M=a_M/(2GM)$. Because of the factors ${(k/k_M)}^4$ and $\Theta (1-\frac{k}{k_M})$, contributions from
both the super-Hubble perturbations and the sub-Hubble perturbations to $\sigma_M^2$ are suppressed.
Importantly, contrary to $\sigma_M$ in the case of the halo formation, $\sigma_M$ given above is not a
monotonically increasing function of $1/M$.
For instance, for the sharply peaked power spectrum at $k=k_*$;
\be
\label{peaked-spectrum}
{\cal P}_\zeta (k)=
{\cal A} k_* \,\frac{1}{\sqrt{2\pi}\Delta} e^{-\frac{(k-k_*)^2}{2\Delta^2}} ,~~~~~{\cal P}_\zeta =\frac{k^3}{2\pi^2} P_\zeta,
\ee
which in the limit $\Delta \rightarrow0$, gives ${\cal P}_\zeta (k)= {\cal A} k_* \delta (k-k_*)$, and we have
\be
\sigma_M^2=\frac{16{\cal A}}{81} {\left( \frac{M}{M_*} \right)}^2 \Theta (M_*-M),
\ee
where $M_*$ is the mass corresponding to $k_*$.
Fig.~\ref{fig:sigma-2-PBH} plots $\sigma_M^2$ in the present case (${\cal A}=\frac{81}{16}$ is assumed).
We find that $\sigma_M^2$ is sharply peaked at around the typical mass scale $M_*$,
which is expected given that $\sigma_M^2$ roughly measures the amplitude of ${\cal P}_\zeta$ at the mass scale $M$.

As in the case of the halo formation, let us again introduce the parameter $\tau$ 
by $\tau=a_M/GM (\propto 1/M^{1/2}$, where we used $a \propto H^{-1/2}$ during RD era)
and view $\delta ({\bm x}, \tau)$ as a stochastic process obeying the equation (\ref{Langevin-eq}) with the initial condition 
$\delta ({\bm x},0)=0 $ and the two-point function given by
\be
\label{del-del-pbh}
\langle \delta (\tau) \delta (\tau') \rangle =
\frac{16}{81} \int \frac{d{\bm k}}{{(2\pi)}^3} P_\zeta (k)
{\left( \frac{k}{\tau} \right)}^2 {\left( \frac{k}{\tau'} \right)}^2 
\Theta (1-\frac{k}{\tau}) \Theta (1-\frac{k}{\tau'}).
\ee
This is achieved by choosing the two-point function of the noise $\xi$ as
\begin{align}
	\label{PBH-C-term}
	C(\tau,\tau')&=\frac{\partial^2}{\partial \tau \partial \tau'} 
	\langle \delta (\tau) \delta (\tau') \rangle \nonumber \\
	&=\frac{4}{\tau \tau'}\langle \delta (\tau) \delta (\tau') \rangle
	-\frac{32}{81} \left( \frac{\tau{\cal P}_\zeta (\tau)}{\tau'^3} \Theta (\tau'-\tau)
	+(\tau \leftrightarrow \tau') \right)+\frac{16}{81} \frac{{\cal P}_\zeta (\tau)}{\tau} \delta 
	(\tau -\tau').
\end{align}
Thus, contrary to the standard PS case (\ref{PS-C-delta}) for which
the noise $\xi$ at different times are uncorrelated,
there are terms which do not vanish even for $\tau \neq \tau'$ in the present case, meaning that the $\xi$ is colored noise.%\footnote{The noise $\xi$ was treated as white noise in Ref.\cite{Auclair:2020csm}.}.

Now, adopting the simple view that PBH is formed when the density contrast smoothed
over the Hubble horizon exceeds a threshold $\delta_c$,
the probability that a given point ${\bm x}$ is inside the region where the PBH 
with its mass larger than $M$ forms is again given by Eq.~(\ref{fraction:PS-formalism}).
Assuming Gaussianity of $\delta$, the first term on the right-hand side of Eq.~(\ref{fraction:PS-formalism})
in the present case is still the same as Eq.~(\ref{P:first-term}).
On the other hand, because of the presence of the correlated term in Eq.~(\ref{PBH-C-term}),
realizing the trajectories like B' is no longer equally likely compared to realizing
the trajectories like B. 

\begin{figure}[t]
	\begin{center}
		\includegraphics[clip,width=7cm]{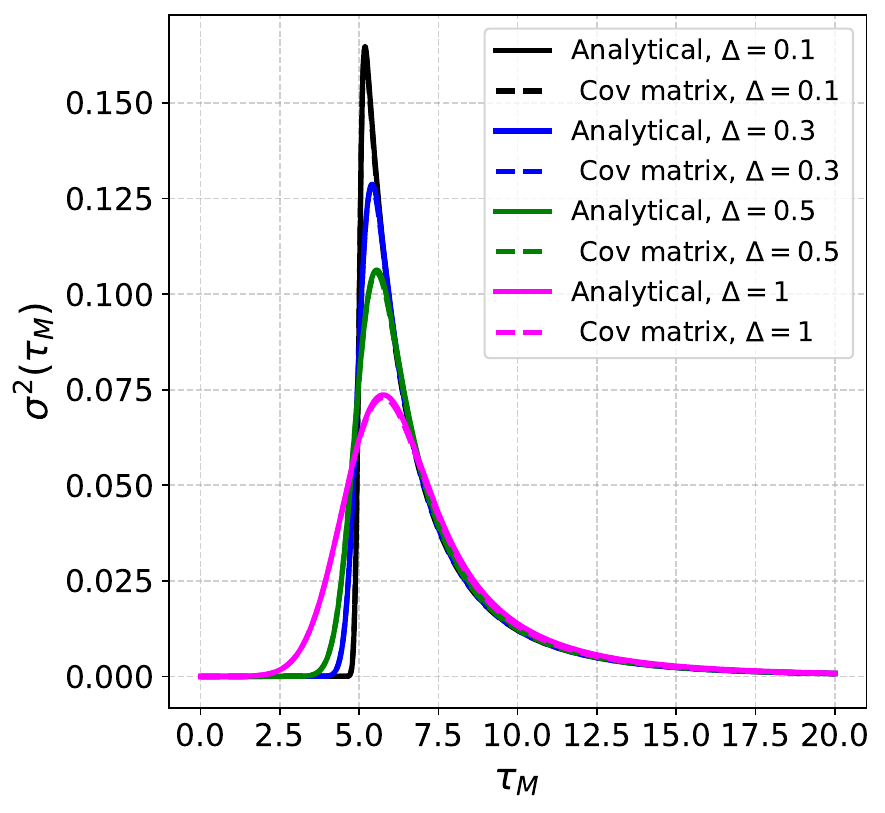}
		\includegraphics[clip,width=7cm]{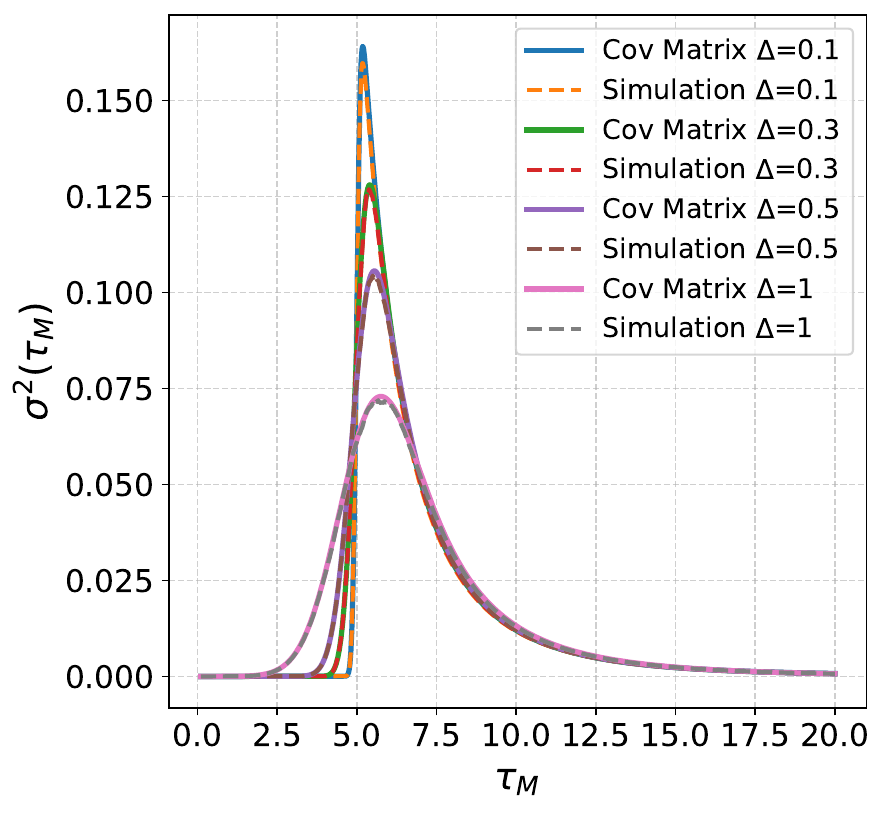}
		\caption{
			Comparison of the analytical and simulated (from  $N=10^4$ realizations) variance $\sigma^2 (\tau_M)$. Left plot: compares analytical variances from Eq.\eqref{exact-del-del-gaussian-approx} with the one computed from the covariance matrix $C(\tau,\tau')$ given in Eq.\eqref{covariance-matrix}, both of these are analytical variances. Right plot: compare the variances obtained from the covariance matrix $C(\tau,\tau')$ given in Eq.\eqref{covariance-matrix} with the simulated data generated by solving the Langevin equation.
			We set $\mathcal{A}=1, \tau_*=5$ and plot for different width $\Delta$.
		}
		\label{fig:pbh-variance-compare-plot}
	\end{center}
\end{figure}

Lastly, using the primordial power spectrum \eqref{peaked-spectrum} in Eq.\eqref{del-del-pbh}, we can obtain the analytical expression 
\begin{align}\label{exact-del-del-gaussian-approx}
	\langle \delta (\tau) \delta (\tau') \rangle =
	\frac{16 \mathcal{A} \tau_*}{81 \, \tau^2 \tau'^2} & \left[
	\frac{\Delta}{\sqrt{2\pi}}
	\left(
	e^{-\frac{\tau_*^2}{2\Delta^2}} (2\Delta^2 + \tau_*^2) 
	- e^{-\frac{(k_{\max}-\tau_*)^2}{2\Delta^2}} 
	(k_{\max}^2 + 2\Delta^2 + k_{\max} \tau_* + \tau_*^2)
	\right) \right.
	\nonumber\\
	&
	\left. + \frac{1}{2} \tau_* (3\Delta^2 + \tau_*^2)
	\left(
	\operatorname{Erf}\!\left(\frac{k_{\max}-\tau_*}{\sqrt{2}\Delta}\right)
	+ \operatorname{Erf}\!\left(\frac{\tau_*}{\sqrt{2}\Delta}\right)
	\right)
	\right]
\end{align}
where $k_{\max} = \min (\tau,\tau')$ and $\operatorname{Erf}$ is the error function. 
Using this, we can obtain the exact expression for the analytical variance
%($\to$ exact analytical expression for the variance)
$\sigma^2({\tau_M}) = \langle \delta (\tau_M)^2 \rangle$ with $k_{\max} = \tau_M$. 
In Figure~\ref{fig:pbh-variance-compare-plot} (left plot), 
we compare the analytical variance computed from Eq.\eqref{exact-del-del-gaussian-approx} with the one calculated from the covariance matrix~\eqref{PBH-C-term} as 
\begin{equation}
	\label{covariance-matrix}
	\sigma^2({\tau_M}) = \int_0^{\tau_M} \int_0^{\tau_M} d\tau d\tau' C(\tau,\tau').
\end{equation}
As we can see, both are consistent with each other.

\begin{figure}[t]
	\begin{center}
		\includegraphics[clip,width=7cm]{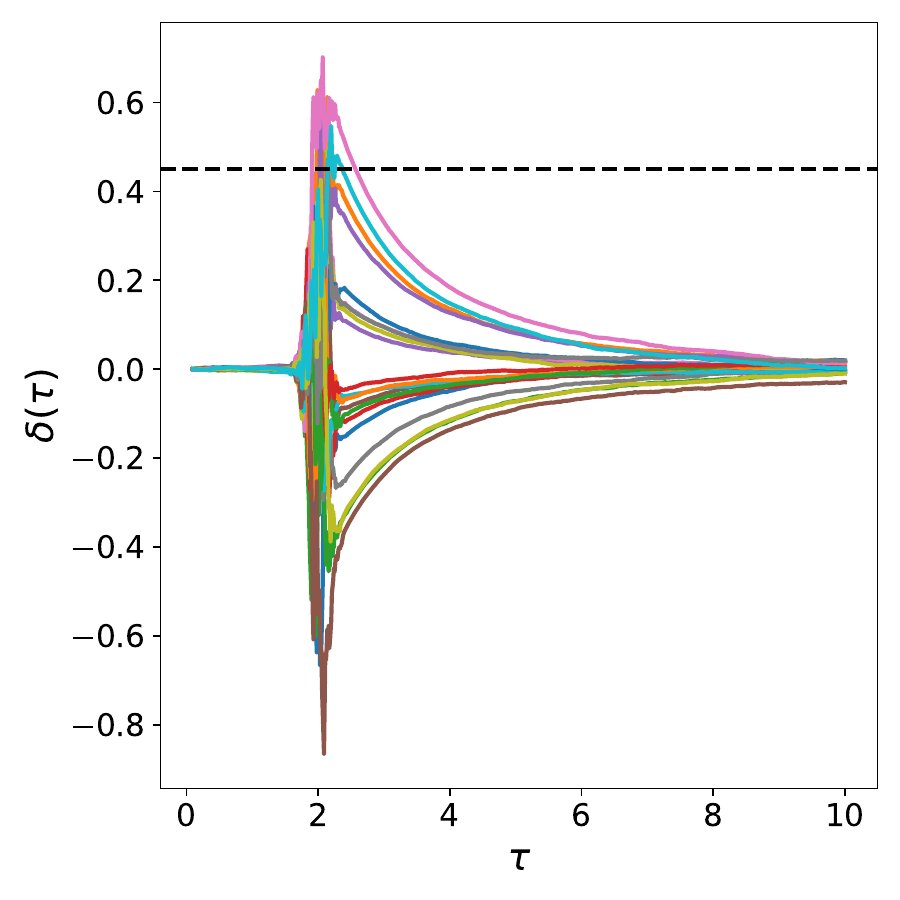}
		\includegraphics[clip,width=7cm]{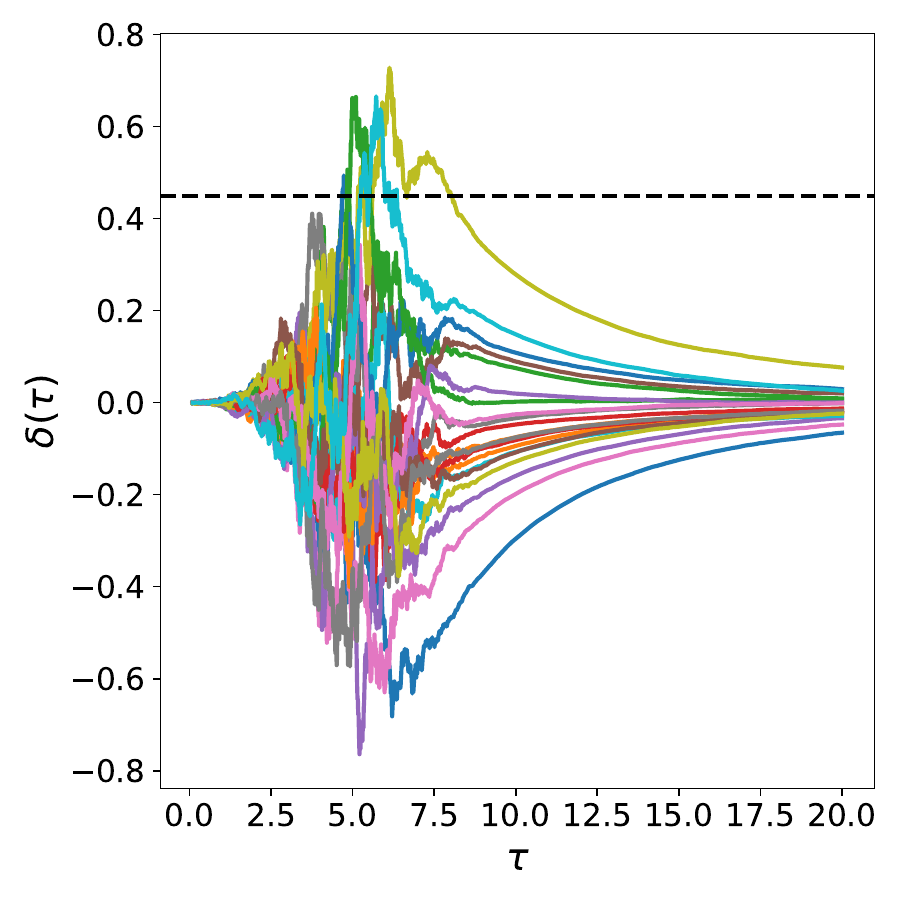}
		\caption{Showing the variation of smoothed density contrast $\delta (\tau )$ (Brownian motion) for the $20$ Langevin trajectories for sharply peaked power spectrum~\eqref{peaked-spectrum} for the PBH formation case where the covariance matrix is given by the correlated noise (colored noise) as in Eq.\eqref{PBH-C-term}. Left plot: narrow peak with $\Delta=0.1$ and $\tau_*=2, \mathcal{A}=1$. Right plot: broad peak with $\Delta=1$ and $\tau_*=5, \mathcal{A}=1$,
			The horizontal dashed line shows the typical value of critical density contrast $\delta_c=0.45$.
		}
		\label{fig:pbh-spiky-power-spectrum}
	\end{center}
\end{figure}

\begin{figure}[t]
	\begin{center}
		\includegraphics[clip,width=7cm]{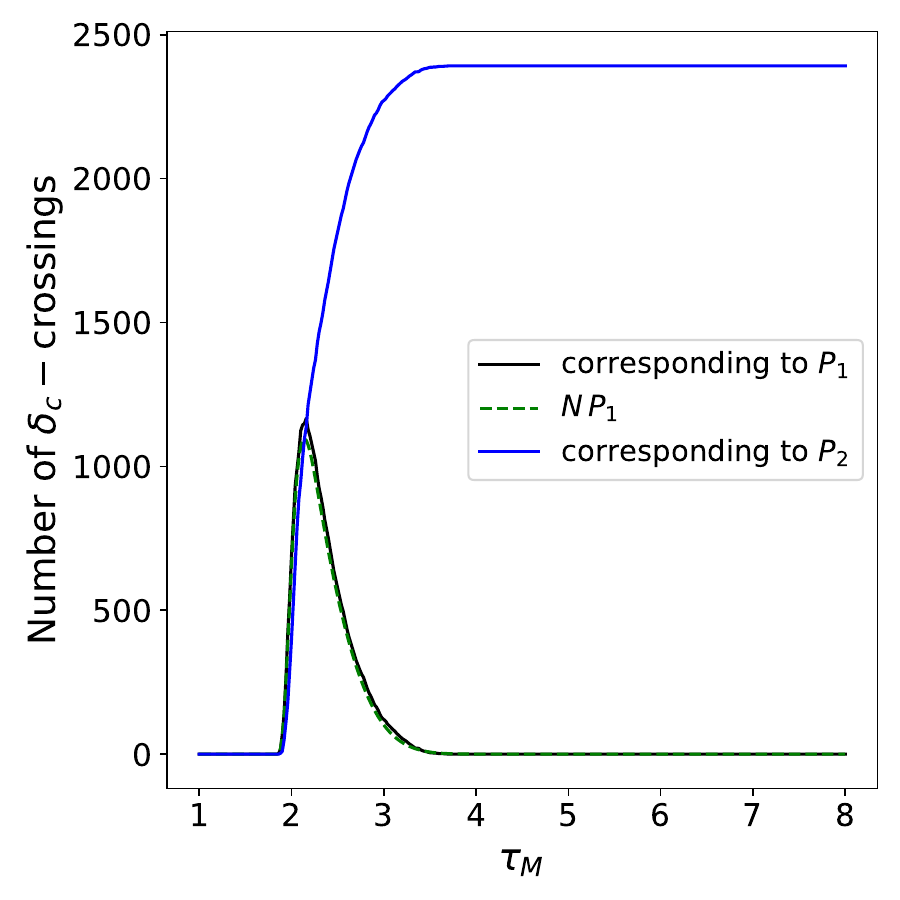}
		\includegraphics[clip,width=7cm]{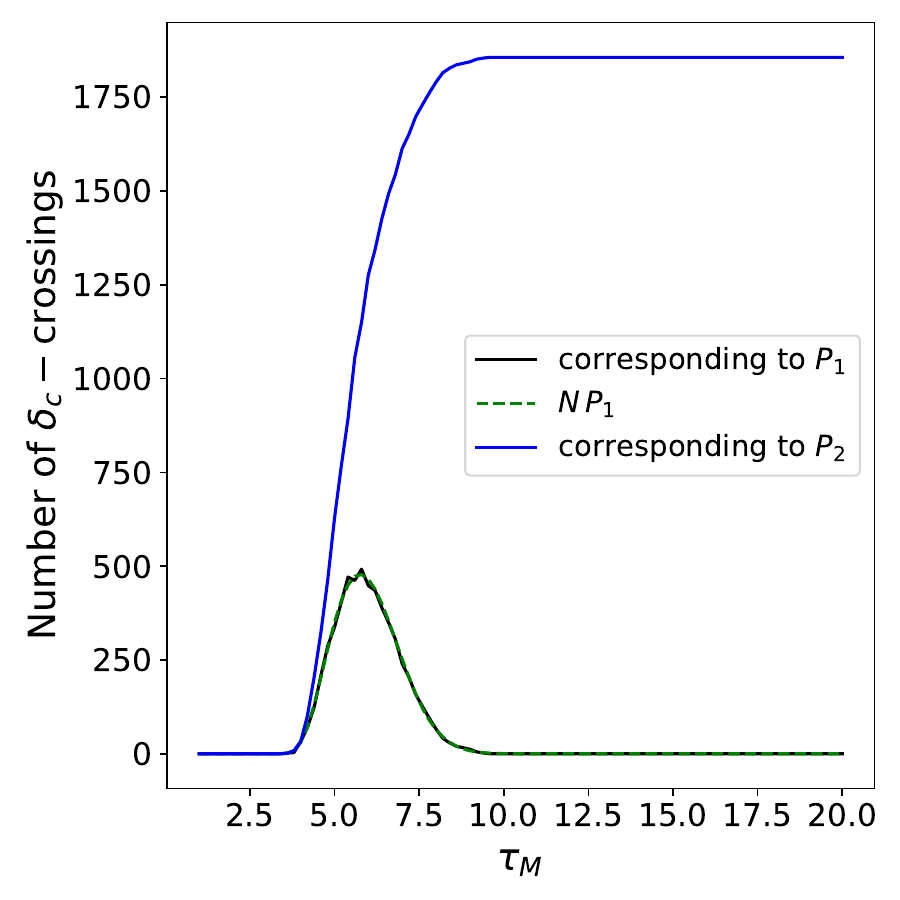}
		\caption{Left panel: The black and blue curves represent the number of trajectories computed from $N=10^4$ realizations, corresponding to $P_1$ and $P_2$, respectively, for sharply peaked power spectrum~\eqref{peaked-spectrum} for $\Delta=0.1,\tau_*=2$ for the PBH formation case where the covariance matrix is given by the correlated noise (colored noise) as in Eq.\eqref{PBH-C-term}.
			The green dotted curve shows $N P_1 (\tau_M)$ computed from Eq.~(\ref{P:first-term}) which is the expected number of trajectories corresponding to $P_1$.
			The agreement between the black curve and the green dotted curve, 
			within statistical uncertainty, indicates that the simulations were performed correctly.
			Right panel: Same as the left panel but for broad power spectrum with $\Delta=1,\tau_*=5$.
		}
		\label{fig:pbh-spiky-power-spectrum-cross-stat}
	\end{center}
\end{figure}

\begin{figure}[t]
	\begin{center}
		\includegraphics[clip,width=7cm]{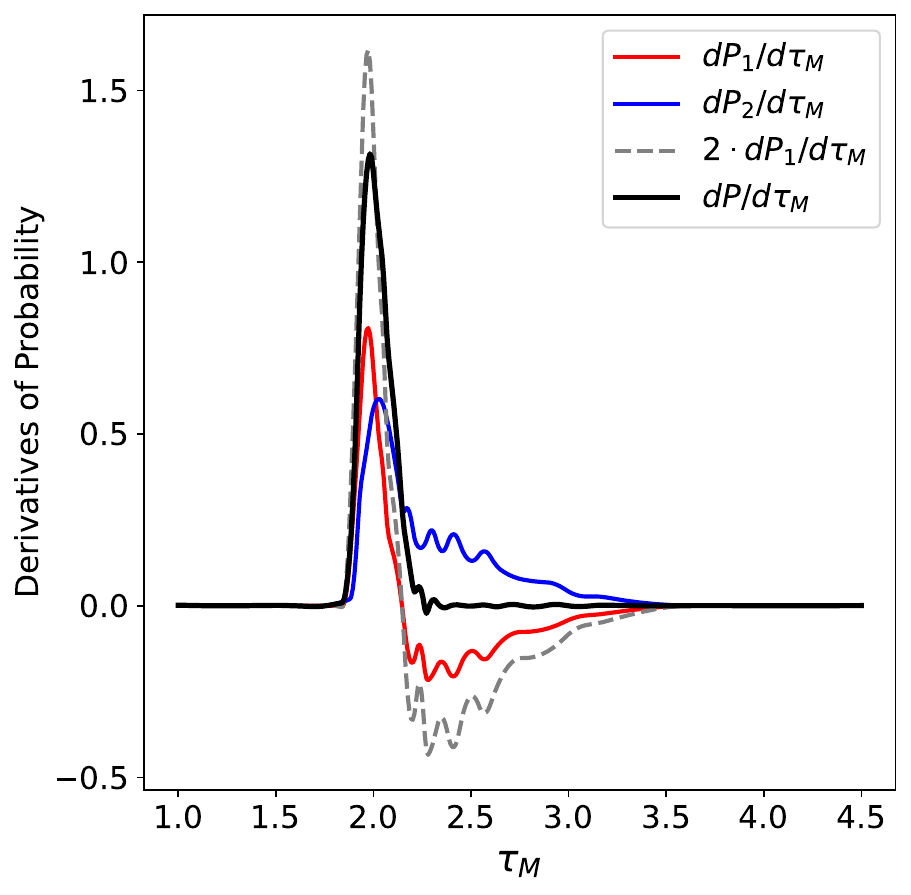}
		\includegraphics[clip,width=7cm]{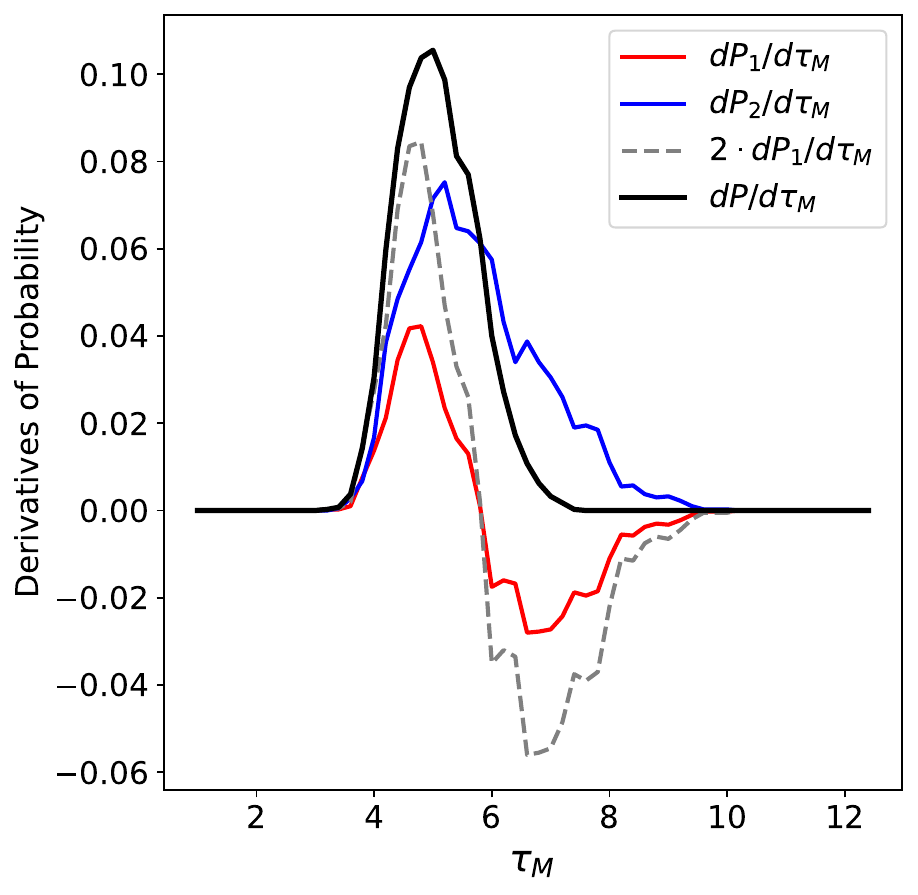}
		\caption{
			Left panel: The derivatives of $P_1 $ and $P_2$ for the number of trajectories computed from $N=10^4$ realizations, for sharply peaked power spectrum~\eqref{peaked-spectrum} for $\Delta=0.1,\tau_*=2$ for the PBH formation case where the covariance matrix is given by the correlated noise (colored noise) as in Eq.\eqref{PBH-C-term}. Right panel: Same as the left but for $\Delta=1,\tau_*=5$.
		}
		\label{fig:pbh-prob-derivatives}
	\end{center}
\end{figure}

\subsection{Numerical simulations}
To investigate the need of the `fudge factor 2' for the PBH formation case, we follow the same analysis as in the previous section. 
We simulate the Langevin equation \eqref{Langevin-eq}, with the covariance matrix~\eqref{PBH-C-term}
with the primordial powers pectrum given by \eqref{peaked-spectrum}. 
However, in this case, the noise $\xi (\tau)$ is correlated at different times, which is called colored noise~\cite{1995-Hanggi.Jung}. 
Generating colored noise is non-trivial; therefore, it is important to first check if the noise $\xi(\tau)$ generated from the covariance matrix~\eqref{PBH-C-term} is correct. 
To do that, we can estimate the variance by using the generated noise $\xi$, which is called the simulated variance, and compare it with the analytical variance obtained from the covariance matrix in 
\eqref{covariance-matrix}. For a sample of $N=10^4$ trajectories, this is shown in Figure \ref{fig:pbh-variance-compare-plot} (right plot), which shows a good match between theoretical and empirical variances. 
This justifies that the number of trajectories, $N=10^4$ are statistically sufficient in our discussion.

Let us now simulate the behaviour of Brownian motion of $20$ Langevin trajectories for narrow peak ($\Delta=0.1$) and broad peak ($\Delta=1$), which is shown in Figure~\ref{fig:pbh-spiky-power-spectrum}. 
As we can see in both cases, the variance goes to zero as $\tau_M \gg \tau_*$.
which is consistent with the analytical result Eq.~(\ref{exact-del-del-gaussian-approx}).
Note that for the case of PBHs formation, narrow peak in the power spectrum produces all of the PBHs with the same masses, i.e., \emph{monochromatic PBH mass function}, while the broad peak lead to the formation of PBHs in a finite mass range, 
which are referred to as \emph{extended PBH mass function}. We will discuss this in detail in the next subsection.\\
To understand the $\delta_c-$crossing statistics of the Langevin trajectories, in Figure~\ref{fig:pbh-spiky-power-spectrum-cross-stat}, we plot a sample of $N=10^4$ Langevin trajectories with varying $\tau_M$ and count the number of paths crossing $\delta_c$ at $\tau_M$ i.e., 
correspoding to the first term $P_1$ in Eq.\eqref{fraction:PS-formalism} and the second term 
$P_2$ in Eq.\eqref{fraction:PS-formalism}. 
In Figure~\ref{fig:pbh-spiky-power-spectrum-cross-stat}, we also show the curve which estimate the number of $\delta_c-$crossing at $\tau_M$ from the theoretical probability $N\,P_1$ given by the Eq.\eqref{P:first-term}. Thus, Figure~\ref{fig:pbh-spiky-power-spectrum-cross-stat} essentially shows plots for $P_1$ and $P_2$ multiplied by $N=10^4$.

We observe that the number count for both narrow width and broad width differs, i.e., $P_1 \neq P_2$. Thus, we can conclude that $P_1 $ and $P_2$ in Eq.\eqref{fraction:PS-formalism}, are not equally probable. Therefore, the `fudge factor 2' in the PBH case would be irrelevant in the sense that $P (>M) = P_1 + P_2 \neq 2\cdot P_1$. We also notice that after $\tau_*$, specifically in the regime $\tau_M \gg \tau_*$, the probability of the second term is greater than the first term in Eq.\eqref{fraction:PS-formalism}. This is because, as we can see from Figure.\eqref{fig:pbh-spiky-power-spectrum}, the variance vanishes as $\tau_M \gg \tau_*$, hence there will not be any trajectory above the $\delta_c$ in that regime. Thus, we only have trajectories corresponding to the second term. This behaviour is solely due to the correlated noise structure of the $C (\tau,\tau')$ term in Eq.\eqref{PBH-C-term}. This is in contrast with the usual halo formation case, discussed in the previous section.

\subsection{Computation of mass function}
The probability that a given point ${\bm x}$ is in a collapse object whose mass is greater than $M$ is given by $P(>M)$ in Eq.\eqref{fraction:PS-formalism}. For two 
arbitrary masses with $M_s < M_L$, the following relation $P (>M_s) = P (>M_L) + P(M\in(M_s,M_L))$ holds, where $P(M\in(M_s,M_L))$ is the probability of being in an object within the mass range $(M_s,M_L)$ and therefore a positive quantity. This suggests a hierarchy between the probabilities of the smaller ($M_s$) and larger ($M_L$) masses, which follows $P (>M_s) > P (>M_L)$. 
Therefore, we can conclude that in general $P(>M)$ is a monotonically decreasing function of $M$, i.e., $\frac{dP(>M)}{dM} < 0$. 
This allows us to obtain the probability that a given point would be in the collapsed object in the infinitesimal mass range
$(M,M+dM)$ as (irrespective of the halo or PBH case) 
\begin{align}\label{mass-fun-general}
	P(>M) - P(>M+dM) = - \frac{\partial P(>M)}{\partial M} dM~.
\end{align}
Using the mass function $\frac{dn (M)}{dM}$, the number density of the collapsed objects in the infinitesimal mass range $(M,M+dM)$ is given by
$\frac{dn (M)}{dM}dM$ (definition of the mass function).
Thus, up to a positive numerical coefficient which depends on $M$,
the mass function is equal to $ \frac{\partial P(>M)}{\partial \tau_M}$ 
(note that $\frac{dM}{d\tau_M}$ is negative).
As our primary interest in the mass function is the dependence of $P$ on $\tau_M$,
we concentrate on this component.
In Figure~\ref{fig:pbh-prob-derivatives}, we plot $\frac{dP}{d\tau_M}$ and also $\frac{dP_1}{d\tau_M}, \frac{dP_2}{d\tau_M}$.
As we can see, $\frac{dP}{d\tau_M}$ remains positive for the entire range of $\tau_M$ 
apart from the right-side tail of the spike in the left plot 
where the corresponding curve (black) exhibits oscillatory feature.
The small wiggles are also present for the curves corresponding to $\frac{dP_1}{d\tau_M}$ and 
$ \frac{dP_2}{d\tau_M}$.
These wiggles arise due to our interpolation scheme, but the overall shape is independent of the scheme and represents genuine physical features.
Apart from the small wiggles, the fact that the numerically computed $\frac{dP}{d\tau_M}$ 
remains positive suggests that the simulations have been performed with reasonable accuracy.
Interestingly, our simulations show that the positivity does not hold for $\frac{dP_1}{d\tau_M}$,
demonstrating that inclusion of $\frac{dP_2}{d\tau_M}$ is essential to correctly evaluate
the mass function.
This result shows that the mass function must be computed with respect to $\frac{dP}{dM}$, not with $\frac{dP_1}{dM}$ as often used in the literature~\cite{Kim:1996hr,Sureda:2020vgi}. 

Let us briefly discuss the calculations followed in the literature to calculate the PBH mass function based on the formula provided by the original PS formalism for halos, which assumes $P_1=P_2$, thus $\frac{dP}{dM}=2\frac{dP_1}{dM}$ with a `fudge factor 2'. 
The mass function is given by\cite{Kim:1996hr,Sureda:2020vgi}
\begin{align}\label{pbh-mass-function}
	\frac{dn (M)}{dM} = f(\nu)\frac{\rho_{m,0}}{M} \frac{d\nu}{dM} = -f(\nu)\frac{\rho_{m,0}}{M} \frac{\delta_c}{\sigma(M)^2} \frac{d\sigma(M)}{dM}
\end{align}
where $\nu(M) = \frac{\delta_c}{\sigma(M)}$, and 
\begin{align}
	f(\nu) = \frac{2}{\sqrt{2\pi}} \exp\left( -\frac{1}{2} \nu (M)^2\right) ~~.
\end{align}
Now, considering the first term in Eq.\eqref{fraction:PS-formalism}, 
\begin{equation}\label{P1-term}
	P_1 =\int_{\delta_c}^\infty \frac{1}{\sqrt{2\pi} \sigma (M)}
	\exp \left( -\frac{\delta^2}{2\sigma^2 (M)} \right) d\delta = \frac{1}{2} {\rm Erfc}\left( \frac{\delta_c}{2\sigma(M)} \right) 
\end{equation}
where ${\rm Erfc}$ is the complementary error function, the derivative of $P_1$, provides a relation to obtain the PS mass function defined in Eq.\eqref{pbh-mass-function} as
\begin{align}\label{pbh-mass-function-P1-rel}
	\frac{dn (M)}{dM} = -\frac{\rho_{m,0}}{M} \left( 2 \frac{d P_1}{dM} \right)
\end{align}
where the term $2dP_1/dM$ on the RHS corresponds to the dashed curves (for $2dP_1/d\tau_M$) shown in Figure~\ref{fig:pbh-prob-derivatives}. As we have already discussed, the positive definiteness of the mass function given by \eqref{pbh-mass-function-P1-rel} does not hold true, and the calculations with $\frac{dP_1}{dM}$ would give incorrect results. 
This is one of the key result of this work, and to the best of our knowledge, it has not been pointed out in the literature. 
Although Ref.~\cite{Sureda:2020vgi} suggests that negativity of the mass function
computed from $\frac{dP_1}{dM}$ has been addressed in the literature (see the references therein), 
we could not find any definitive discussion in the cited works. 
Notably, in these works the negative mass function appears for spectral index $n_s <1$, 
but this case does not seem to be explicitly considered.

Before closing this section, we would like to remark the computation of PBH abundance, which is often defined as\cite{Garcia-Bellido:1996mdl,Yokoyama:1998pt,Young:2014ana,Garcia-Bellido:2017mdw,Inomata:2018cht,2018-Sasaki.etal-CQG,Bhaumik:2019tvl,Braglia:2020eai,Gow:2020bzo,2020-Carr.etal-Rept.Prog.Phys,Kawasaki:2012wr} 
\begin{align}
	\beta (M) =\int_{\delta_c}^\infty \frac{1}{\sqrt{2\pi} \sigma (M)}
	\exp \left( -\frac{\delta^2}{2\sigma^2 (M)} \right) d\delta = \frac{1}{2} {\rm Erfc}\left( \frac{\delta_c}{2\sigma(M)} \right)
\end{align}
where often the `fudge factor $2$' is multiplied or ignored. However, From Eq.\eqref{P1-term}, we can see that, above equation corresponds to $\beta (M) = P_1(M)$ and with the `fudge factor $2$' (i.e., assuming incorrectly $P_1 = P_2$) corresponds to $\beta (M) = 2 \cdot P_1(M)$. 

\section{Conclusion}

PBHs are a unique probe of the early Universe and are considered as potential dark matter candidate. The formation of PBHs have been a subject of serious investigations due to various uncertainties related to the methods for computing the abundance. One popular method is the PS formalism, which was originally proposed to compute the mass distribution of the virialized halos. The original PS formalism for the halo formation case suffers from the cloud-in-cloud problem and an ad hoc multiplicative `fudge factor 2’ was introduced without rigorous justification. 
The excursion set theory approach proposed by Bond et al.~\cite{Bond:1990iw}, provides a rigorous framework to systematically show the justification of multiplication by `fudge factor 2’. However, the naive application of the PS formalism to PBHs created the confusion of considering or omitting the multiplication by `fudge factor 2' in the calculations of PBH abundance. 
The mathematically rigorous justification for either of the choices is still missing in the literature. 

In this work, we studied the formulation of PS formalism for PBHs forming during RD epoch within the excursion set formalism, where the smoothed density contrast at a fixed spatial point $\delta(\tau_M)$ evolves stochastically (random walk) with the smoothing scale ($\tau_M^{-1}$), and the collapse is identified with the first crossing of a critical threshold $\delta_c$. 
Unlike the case of the halo formation, we have shown that for any primordial power spectrum,
the stochastic evolution of $\delta (\tau_M)$ is non-Markovian for the case of PBHs, 
even when the sharp-k filter is adopted as the Window function. To investigate this in detail, we performed numerical simulations of the evolution of $\delta (\tau_M)$ by solving the Langevin equation, and found that the non-Markovian nature for the PBHs case is reflected in the region $\tau_M \gg \tau_* $ with the variance, $\sigma^2 (\tau_M)\rightarrow0$. Furthermore, a consistent comparison of the probability of the first $\delta_c$-crossing at a given time $\tau_M$ (i.e., $P_1$) with that of a crossing occurring prior to 
$\tau_M$ (i.e., $P_2$) reveals that $P_1 \neq P_2$. This represents a characteristic feature of PBHs, in contrast to halos, for which $P_1 = P_2$ holds.
Thus, the probability that a given point is in a part of PBH with mass $>M$ follows $P(>M)= P_1 + P_2 \neq 2 P_1$. 
Since the computation of $P_2$ is highly nontrivial and requires solving the Langevin equation numerically, it becomes difficult to provide an analytical formula to calculate $P(>M)$ for the PBHs case.

Our numerical analysis revealed that the contribution from $P_1$ to the PBH mass function
can be negative for some mass range.
Although this negativeness is pointed out in the literature, the correct prescription to
derive the physically sensible mass function has remained missing.
We showed that the contribution from $P_2$, which has been overlooked in the literature, 
is essential to obtain a correct (hence positive) mass function.
Our results clarify the ambiguity on the use of the fudge factor and establish a robust theoretical foundation for PBH abundance calculations.

\section*{Acknowledgements} 
\label{sec:acknowledgements}
The work of A.K. was supported by the Japan Society for the Promotion of Science (JSPS) as part of the JSPS Postdoctoral Program (Grant Number: 25KF0107). 
T.S. gratefully acknowledges support from JSPS KAKENHI grant (Grant Number JP23K03411). A.K. would like to thank Joseph P Johnson for the discussion. We also thank Pierre Auclair and Vincent Vennin for insightful discussion on their work.

\appendix

	\section{Details on noise generation}
	In this section, we briefly describe how to generate colored (i.e., correlated) noise for the numerical simulation of the Langevin equation~\ref{Langevin-eq} when the noise is characterized by a non-trivial covariance matrix $C_{ij}$ (see e.g.~\cite{1995-Hanggi.Jung}).
	Let us begin with the simpler case of uncorrelated (white) noise. In discrete (effective) time, the noise $\xi(\tau_i)$ is characterized by the two-point correlation function
	\begin{align}
		\langle \xi(\tau_i)\,\xi(\tau_j) \rangle = \sigma^2 \delta_{ij},
	\end{align}
	where $\sigma^2$ is the variance and $\delta_{ij}$ is the Kronecker delta. In this case, the noise values at different time steps are statistically independent. 
	Numerically, white noise is generated by drawing independent Gaussian random variables
	$\xi_i \sim \mathcal{N}(0,\sigma^2)$ from a normal distribution with zero mean and variance $\sigma^2$ at each time step. When such white noise drives the Langevin equation $\frac{d\delta}{d\tau} = \xi(\tau)$, the resulting process $\delta(\tau)$ corresponds to a random walk (Wiener process), i.e., Brownian motion. Because the increments are independent and depend only on the current state, the process is \emph{Markovian}, meaning that its future evolution depends solely on the present value and not on its history.

For colored noise, which is characterized by a non-diagonal covariance matrix,
	\begin{align}
		\langle \xi(\tau_i)\,\xi(\tau_j) \rangle = C_{ij},
	\end{align}
	where $C_{ij} \neq 0$ for $i \neq j$. The noise values at different time steps are therefore correlated. In this case, the noise vector must be drawn from a multivariate normal distribution with covariance matrix $C$.
	Our aim is to generate a single realization of the random vector $\boldsymbol{\xi} = \left( \xi_1, \xi_2, \dots, \xi_n \right)^T$
	%
	%\begin{align}
	%\boldsymbol{\xi} = \left( \xi_1, \xi_2, \dots, \xi_n \right)^T,
	%\end{align}
	%
	such that $\boldsymbol{\xi} \sim \mathcal{N}(\mathbf{0}, C)$.
	%
	%\begin{align}
	%\boldsymbol{\xi} \sim \mathcal{N}(\mathbf{0}, C)~.%,
	%\end{align}
	%
	%i.e., a multivariate Gaussian distribution with zero mean and covariance matrix $C$. 
	To do that, we first, we generate a vector of independent standard normal random variables,
	\begin{align}
		\mathbf{z} = (z_1, z_2, \dots, z_n)^T,
		\qquad
		\mathbf{z} \sim \mathcal{N}(\mathbf{0}, \mathbb{I}),
	\end{align}
	where $\mathbb{I}$ is the identity matrix and all components $z_i$ are independent with unit variance. The key step is to perform the Cholesky decomposition of the covariance matrix $C$, which is defined for symmetric or Hermitian positive-definite matrices. The Cholesky factorization of $C$ is given as
	\begin{align}
		C = L L^T,
	\end{align}
	where Cholesky factor $L$ is a lower-triangular matrix. This decomposition exists provided that $C$ is symmetric and positive definite. Using the Cholesky factor $L$, we can construct the correlated noise vector as
	\begin{align}
		\boldsymbol{\xi} = L\,\mathbf{z}.
	\end{align}
	It is straightforward to verify that this construction yields the desired covariance matrix
	\begin{align}
		\langle \boldsymbol{\xi}\boldsymbol{\xi}^T \rangle
		= L \langle \mathbf{z}\mathbf{z}^T \rangle L^T
		= L \mathbb{I} L^T
		= C.
	\end{align}
	Therefore, $\boldsymbol{\xi}$ is a realization of a multivariate Gaussian random vector with the given covariance matrix $C_{ij}$. This procedure provides a consistent and numerically stable method to generate colored noise for Langevin simulations.

In contrast to the white-noise case, where $C_{ij} \propto \delta_{ij}$ and successive increments are statistically independent, here $C_{ij} \neq 0$ for $i \neq j$. The noise values at different time steps are therefore correlated, implying that the increments are no longer independent. As a consequence, a Langevin equation driven by such colored noise generally defines a non-Markovian process, since the future evolution depends not only on the present state but also on its history through the temporal correlations encoded in $C_{ij}$.

%\bibliographystyle{apsrev4-1}
\input{PRD-Ver-2.bbl}
%\bibliography{ref}
\end{document}

%% file: PRD-Ver-2.bbl
%merlin.mbs apsrev4-1.bst 2010-07-25 4.21a (PWD, AO, DPC) hacked
%Control: key (0)
%Control: author (72) initials jnrlst
%Control: editor formatted (1) identically to author
%Control: production of article title (-1) disabled
%Control: page (0) single
%Control: year (1) truncated
%Control: production of eprint (0) enabled
%